\documentclass[11pt,a4paper]{article} 
\usepackage{jheppub}

\usepackage{slashed}
\usepackage{youngtab}

\newcommand{\beq}{\begin{equation}}
\newcommand{\eeq}{\end{equation}}

\newcommand{\beqa}{\begin{eqnarray}}
\newcommand{\eeqa}{\end{eqnarray}}

\newcommand{\bn}{\begin{equation}}
\newcommand{\en}{\end{equation}}
\newcommand{\by}{\begin{eqnarray}}
\newcommand{\ey}{\end{eqnarray}}
\newcommand{\al}{\alpha}
\newcommand{\be}{\beta}
\newcommand{\ga}{\gamma}
\newcommand{\Ga}{\Gamma}
\newcommand{\de}{\delta}
\newcommand{\ep}{\epsilon}
\newcommand{\om}{\omega}
\newcommand{\si}{\sigma}
\newcommand{\thalf}{\tfrac{1}{2}}
\newcommand{\pd}{\partial}

\newcommand\fverb{\setbox\fverbbox=\hbox\bgroup\verb}
\newcommand\fverbdo{\egroup\medskip\noindent%
            \fbox{\unhbox\fverbbox}\ }
\newcommand\fverbit{\egroup\item[\fbox{\unhbox\fverbbox}]}
\newbox\fverbbox

\newcommand{\nablaslash}{\not{\hbox{\kern-3pt $\nabla$}}}

\title{Critical solutions of topologically gauged  ${\mathcal N}=8$ CFTs in three dimensions}
\author{Bengt E.W.~Nilsson%\\
\\
Fundamental Physics\\
Chalmers University of Technology\\
SE-412 96 G\"oteborg, Sweden\\

{\tt {\footnotesize  tfebn@chalmers.se}}}

\abstract{ In this paper we discuss some special (critical) background solutions that arise in  topological gauged  ${\mathcal N}=8$  
three-dimensional CFTs with $SO(N)$  gauge group. Depending on how many scalar fields are given a VEV the theory has background solutions
for certain values of $\mu l$, where $\mu$ and $l$ are parameters in the $TMG$ Lagrangian. 
Apart from  Minkowski, chiral round $AdS_3$ and 
null-warped $AdS_3$ (or Schr\"odinger($z=2$)) we identify also  a  more exotic  solution   recently found  in $TMG$ by Ertl, Grumiller and Johansson. 
We also discuss the spectrum, symmetry breaking pattern and 
the supermultiplet structure in the various backgrounds and argue that some properties are due to their  common origin in a conformal phase. Some of the  scalar fields, including all higgsed ones,  turn out to satisfy  three-dimensional  field equations similar to those of the  singleton. Finally, we note that  topologically gauged ${\mathcal N}=6$
ABJ(M) theories have a  similar, but more restricted, set of background   solutions.}
 
\keywords{String theory, M-theory, Branes, Chern-Simons theory, AdS/CFT}

\toccontinuoustrue
\begin{document}
\maketitle

%\setcounter{page}{2}

%%%%%%%%%%%%%%%%%%%%%%%%%%%%%%%%%%%%%%%%%%%%%%%%%%%%%%%%%%%%

\section{Introduction}

%%%%%%%%%%%%%%%%%%%%%%%%%%%%%%%%%%%%%%%%%%%%%%%%%%%%%%%%%%%%

The purpose of this paper is to discuss some of the background solutions   in topologically gauged CFTs in $2+1$ dimensions with  
$\mathcal N=8$ supersymmetry and an arbitrary $SO(N)$ gauge group \cite{Gran:2008qx,Gran:2012mg} and to
point out  some of their properties relevant in this context. Apart from Minkowski, and  well-known geometries like  round $AdS_3$ and null-warped $AdS_3$ found already in 
 \cite{Gran:2012mg}, we here identify a new 
more exotic one belonging to a different category of solutions as will be explained below. The main point of this paper is to argue that only very special solutions
in topological massive gravity (TMG) 
will appear due to the connection to the unbroken superconformal phase of the theory.

Topological gauged CFT refers in general to superconformal  Chern-Simons(CS)/matter field theories in three dimensions whose global symmetries have been gauged by coupling the theory to conformal  supergravity. In three dimensions conformal supergravity is governed by gravitational CS terms \cite{vanNieuwenhuizen:1985cx,Lindstrom:1989eg} and is therefore topological in nature.
Topologically gauged CFTs of this kind were first discussed in \cite{Gran:2008qx} where the gauging was applied to the ordinary $\mathcal N=8$ BLG  theory 
\cite{Bagger:2006sk, Bagger:2007jr, Gustavsson:2007vu}. For the $\mathcal N=6$ ABJ(M) theories \cite{Aharony:2008ug,Aharony:2008gk} the same type of construction was obtained  shortly afterwards  in \cite{Chu:2009gi} where a new potential for the scalar fields was found as we will have reason to briefly discuss later. Entirely new theories with 
local $\mathcal N=8$ conformal supersymmetry, $SO(N)$ gauge groups for any $N$ and new scalar potential terms were subsequently 
discovered in \cite{Gran:2012mg} which also completed the task, set by the authors of \cite{Gran:2008qx}, of gauging the BLG theory.
 The topological  properties of the gravitational sector of the theory  are important for what kind of degrees of freedom it describes. Thus, one of our goals will  be to initiate an analysis of the spectrum in the different broken phases of the gauged theory with $SO(N)$ gauge symmetry. The higgsing that turns the CS gauge fields into massive vector fields will be discussed in detail, and we will present some exact formulae for the interactions with the remaining scalar fields. We will also note that
 the higgsed scalars satisfy the singleton field equation.%Recall that the construction in  \cite{Gran:2008qx,Gran:2012mg} of the  topological gauged version of the  BLG theory led to a  somewhat unexpected new result: the allowed gauge group is initially
%$SO(4)$ but when turning off the original  BLG gauge interactions (by for instance taking the BLG level parameter to infinity) the gauge group can be generalized to
%$SO(N)$ for any integer $N$. Since these new  theories can be viewed as a topological gauging of the free ${\mathcal N}=8$ superconformal matter theory (without any gauge sector) this 
%has rather little to do with what is normally called "the BLG theory" we will in the following refer to all of them as just "topologically gauged $N=8$ theories" and use "BLG"  only when  appropriate. 
%The purpose of this section is to first review the results of the topological gauging of the BLG theory presented 

The construction of the topological gauged BLG theory was started in \cite{Gran:2008qx} and completed in \cite{Gran:2012mg} where it was also found that if one turns off the BLG interactions it becomes possible to  generalize the   gauge symmetry from $SO(4)$ to $SO(N)$ for any $N$. This was shown using three different methods, one of them being the Noether method.\footnote{The other two methods used in  \cite{Gran:2012mg} are the on-shell superalgebra method and superspace. In that work the superspace  method was finally successfully applied to this problem which has a number of special features that make the analysis more complicated than for Poincar\'e supergravity theories, see, e.g.,  \cite{Howe:1995zm, Cederwall:2011pu}.} Since no  details of the  derivation of the potential using  the Noether method were  given in \cite{Gran:2012mg}
we present some of the  details  in the appendix, restricting ourselves to the $SO(N)$  theory which starts from the free matter theory. We stop the presentation at the point where we can deduce the new potential terms. The appendix also  discusses  the $SO(N)$ gauge field and presents a more direct argument for the normalization of the $SO(N)$ CS term than that given in  \cite{Gran:2012mg}.
The reader may consult \cite{Gran:2012mg} for  the complete arguments showing the existence 
of these  ${\mathcal N}=8$ topologically gauged $SO(N)$ theories.

%%%%%%%%%%%%%%%%%%%%%%%%%%%%%%%%%%%%%%%%%%%%%%%

%\subsection{Review}

%%%%%%%%%%%%%%%%%%%%%%%%%%%%%%%%%%%%%%%%%%%%%%%%

Before we turn to the theory with ${\mathcal N}=8$ let us very briefly review the situation for ${\mathcal N}=6$. The topologically gauged ABJ(M) theories were obtained in \cite {Chu:2009gi} and discussed further in \cite{Chu:2010fk} (see also \cite{Gran:2012mg}). 
Apart from the superconformal gravity sector and a standard ABJ(M) theory it contains  a new $U_R(1)$ CS gauge field and a number of new interaction terms. In particular one finds a new scalar potential and the expected conformal coupling term $-\tfrac{1}{8}|Z|^2R$ between the curvature scalar and two scalar fields $Z^A_a$ which are complex in this case: lower case indices are three-algebra and upper case fundamental $SU(4)$ R-symmetry indices (for details, see \cite {Chu:2009gi}). The potential is then found to  consist of the original 
 (single trace ($st$)) term
 \begin{equation}
V^{(st)}_{ABJ(M)}=\tfrac{2}{3}|\Upsilon^{CD}{}_{Bd}|^2,\,\,
\Upsilon^{CD}{}_{Bd}= \lambda f^{ab}{}_{cd}Z_a^CZ_b^D\bar Z^c_B+\lambda f^{ab}{}_{cd}\delta^{[C}_BZ^{D]}_aZ^E_b\bar Z^c_E \,,
\end{equation}
plus the following new terms:  with one structure constant (double trace ($dt$))
\begin{equation}
V_{new}^{(dt)}=-\tfrac{1}{8} g\lambda f^{ab}{}_{cd}|Z|^2Z^C_aZ^D_b\bar Z^c_CZ^d_D-\tfrac{1}{2}g\lambda f^{ab}{}_{cd}Z^B_aZ^C_b(Z^D_e\bar Z_B^e)\bar Z_C^c\bar Z_D^d \,,
\end{equation}
and without structure constant (triple trace ($tt$))
\begin{equation}
V_{new}^{(tt)}=-g^2(\tfrac{5}{12\cdot 64}(|Z|^2)^3-\tfrac{1}{32}|Z|^2|Z|^4+\tfrac{1}{48}|Z|^6)\,,
\end{equation}
where $\lambda=\tfrac{2\pi}{k}$  ($k$ is the  CS level)  and  $g$ the gravitational coupling constant.

We can now break the conformal symmetries by introducing a real VEV $v$ for one of the scalar fields $Z^a_A$ \cite {Chu:2009gi} and consider the following terms in the lagrangian\footnote{The coupling constant $g$ was later introduced in \cite{Chu:2010fk} but is not really crucial for the argument.}:
\beq
L(v)=-\tfrac{1}{g}L_{CS(\om)}-\frac{v^2}{8}eR-eV(v),
\eeq
where only the triple trace terms contribute to the VEV of the potential $V(v)$. 
By comparing to the TMG Lagrangian discussed in the context of chiral gravity by the authors of \cite{Li:2008dq} (but with an opposite sign in front of  the whole Lagrangian) we find that their parameters can be expressed in terms of the ours, $v$ and $ g$, as follows ($\Lambda=-\tfrac{1}{l^2}$)
\beq
\mu=\tfrac{g}{\kappa^2},\,\, \kappa^2=\tfrac{8}{v^2},\,\,\,\tfrac{1}{2l^2}=\tfrac{g^2v^4}{128},
\eeq
which shows that the chosen VEV produces a theory that sits exactly at the chiral point:
\beq
\mu l=1.
\eeq

%After reviewing  the topological gauged  ${\mathcal N}=8$ CFT with gauge group $SO(N)$ below, 
Below we will repeat the above search for a critical $AdS_3$  solution in the ${\mathcal N}=8$ case.
We will  find that this does not work unless  we generalize the VEV to several scalar fields, a fact first observed in \cite{Gran:2012mg}. This step will generate a set of solutions which will be elaborated upon in section 2. In section 3 we discuss the spectrum and supersymmetry in the various backgrounds. Here relations to $AdS_3$ singletons seem to appear. A few comments are collected in section 4 and some computational details of the Noether construction of the potential can be found in the appendix.

   \section{Field equations and  background solutions}
   
    %%%%%%%%%%%%%%%%%%%%%%%%%%%%%%%%%%%%%%%%%%%

%Since the topologically gauged BLG theory seems to generate a scalar potential that  end up at the AdS  chiral point or null warped chiral AdS, one needs to understand
%if there is a massive (positive energy) graviton or not. There are no propagating gravity modes before the higgsing and hence there should be no
%such modes after the higgsing. However, maybe the new gauging involving the non-abelian vector field mix up the modes which however seems very unlikely
%due to the fact that non-abelian gauge modes can not mix with gravity modes without gauge indices! But there could be new abelian spin fields that could mix! 
%There are in fact no modes that can naturally mix with 
%the gravity modes except possibly the R-symmetry modes but these are also not there in the conformal phase. 

In this section we will find and discuss a number of background solutions. Two of  these were briefly mentioned  in \cite{Gran:2012mg} and  are known to be in some sense  (see below) critical. Here we will also identify a new solution  that is unfortunately  less well understood.
Supersymmetry and other properties of these backgrounds will be discussed in the following section. 

%%%%%%%%%%%%%%%%%%%%%%%%%%%%%%%%%%%%%%%%%%%%%%%%%%%%%%%

\subsection{The bosonic part of the lagrangian with $SO(N)$ gauge symmetry}

%%%%%%%%%%%%%%%%%%%%%%%%%%%%%%%%%%%%%%%%%%%%%%%%%%%%%%%

%The lagrangian is  the sum of the following terms (where the different factors in front of the CS terms should be noted)
%\begin{eqnarray}
%L_{SO(N)}^{kinetic}&=&-\frac{1}{2g}\epsilon^{\mu\nu\rho}
%(\omega_{\mu}{}^{\al\be}\partial_{\nu}\omega_{\rho\al\be}-
%\frac{2}{3}Tr_{\al}\omega_{\mu}\omega_{\nu}\omega_{\rho})
%+\frac{1}{g}\epsilon^{\mu\nu\rho}
%(B_{\mu}^{ ij}\partial_{\nu}B_{\rho}^{ij}-\frac{2}{3}Tr_iB_{\mu}B_{\nu}B_{\rho})\notag\\[1mm]
%&&
%%&&-i e^{-1} \epsilon^{\alpha\mu\nu}\epsilon^{\beta\rho\sigma}(\tilde
%%D_{\mu}\bar{\chi}_{\nu}\gamma_{\beta}\gamma_{\alpha}\tilde
%%D_{\rho}\chi_{\sigma}))
%%\end{eqnarray}
%%\beqa
%%\nn L_{BLG}&=&
%-\tfrac{e}{2}g^{\mu\nu}D_{\mu}X^i_aD_{\nu}X^i_a
%%+\tfrac{i}{2}\bar\psi\ga^{\mu}D_{\mu}\psi
%-\tfrac{2}{g}\epsilon^{\mu\nu\rho}
%(A_{\mu}^{ ab}\partial_{\nu} A_{\rho}^{ab}+
%\tfrac{2}{3} A_{\mu}^{ ab}A_{\nu}^{ac} A_{\rho}^{cb}),
%%+\tfrac{i}{4}\lambda\bar\psi_a\Ga^{ij}\psi_bX^i_cX^j_d\,f^{abcd}
%%-\tfrac{\lambda^2}{12}(f^{abcd}X^i_bX^j_cX^k_d)(f^{aefg}X^i_eX^j_fX^k_g)
%\eeqa
%\beq
%L^{matter}_{gravity}=-\frac{e}{16}X^2\,R,
%\eeq

The bosonic part of the action consists of the following terms \cite{Gran:2012mg} 
\beq
L_{Bos}=-\tfrac{1}{g}L_{CS(\om)}+\tfrac{2}{g} L_{CS(B)} -\tfrac{4}{g} L_{CS(A)}-\tfrac{e}{2} g^{\mu\nu} D_{\mu}X^i_a D_{\nu} X^i_a-\tfrac{e}{16}X^2 R - e V(X),
\eeq
where the various Chern-Simons terms  are given in terms of the conventionally normalized Lagrangians $L_{CS(..)}$: 
\beq
L_{CS(A)}=\frac{1}{2}\ep^{\mu\nu\rho}(A_{\mu}^{ab}\partial_{\nu}A_{\rho}^{ab}+\frac{2}{3}A_{\mu}^{ab} A_{\nu}^{ac} A_{\rho}^{cb}).
\eeq
The conformal coupling $X^2R$ is the $d=3$ version of the general case
\beq
L=-\tfrac{1}{2}(\partial_{\mu}\Phi)^2-\tfrac{d-2}{8(d-1)}\Phi^2R,
\eeq
and the new $SO(N)$ potential, which  is  a special combination of triple trace terms (recall that the BLG structure constants have been set to zero in this $SO(N)$ theory),
%\beq
%L^{new}_{potential}=
%-\tfrac{e\,g^2}{2\cdot 32\cdot 32}((X^2)^3-8(X^2)(X_a^iX_a^j)(X_b^iX_b^j)+16(X_a^iX_a^j)(X_b^jX_b^k)(X_c^kX_c^i))
%\eeq
can be written as a square as follows
\beq
\label{newpotential}
V(X)=\tfrac{g^2}{2\cdot 32\cdot 32}(X^2X^i_a-4X_a^jX_b^jX^i_b   )^2,
\eeq
where the indices $a,b,..$ and $i,j,...$ are vector indices of the gauge group  $SO(N)$ and R-symmetry group $SO(8)$, respectively.
The covariant derivative is  $
D_{\mu}=\partial_{\mu}+ \omega_{\mu}+B_{\mu}+A_{\mu}.
$
See  the appendix for conventions  and \cite{Gran:2012mg}  for  additional details.
%where
%\beq
%\tilde A_{mu}^{ab}:=A_{\mu}^{cd}( \lambda \ep^{cdab}-\tfrac{g}{4}\delta_{ab}^{cd})
%\eeq

We can now vary these terms to get the equations of motion for the bosonic fields which we will  later linearize  to find the spectrum, analyze stability etc.
To properly analyze the issue of stability one needs, in fact, to go beyond the linear level (see, e.g., Maloney et al. \cite{Maloney:2009ck}) but that will not be done in this paper.
%maloney: 0903.4573 and Toy-model -paper:Bergshoeff et al: 1201.0449).

Since  a single scalar VEV $<X>=v$ (for one component $X^1_1$ of $X^i_a$, say) solves the Klein-Gordon equations  we can just insert the VEV into the Lagrangian to analyze which geometries will satisfy the gravitational field (Cotton) equation. To this end we need the background value of the potential:
\beq
V(v)=\tfrac{9g^2v^6}{2\cdot 32\cdot 32}.
\eeq
This is, however, a factor of 9 wrong if we had expected to end up at the chiral point as in the ABJ(M) case \cite{Chu:2009gi}! This is easily seen as follows.
By considering the gravitational CS term, the $X^2R$ term and the potential  evaluated at the VEV we get
\beq
L_{VEV}=-\tfrac{1}{g}L_{CS(\om)}-\tfrac{v^2e}{16}R-eV(v).
\eeq
This may be compared to the action used by Li, Song and Strominger (LSS) \cite{Li:2008dq} in their analysis of  the chiral point\footnote{Note that this is a TMG \cite{Deser:1982sw} type Lagrangian  with signs opposite to those  used by LSS in \cite{Li:2008dq}: the signs used in our paper are dictated by the unitarity of the scalar field sector together with  supersymmetry and can not be changed.
% and would  lead to negative energy black holes in a non-supersymmetric phase. 
 However, even supersymmetric phases may have unitarity problems (appearing here only at the boundary)  as indicated by  the results of \cite{Li:2008dq} and \cite{Becker:2009mk}.
%  It would interesting to see what the effect would be of turning $SO(N)$ into $Sp(N)$ by means of a sign flip $N\rightarrow -N$ as done to obtain a $dS/CFT$ correspondence in \cite{Anninos:2011ui}. In our case we seem, however, to get back $AdS$ since both the Einstein-Hilbert and the cosmological term would  flip sign.
 }:
\beq
L_{LSS}=-\tfrac{1}{\kappa^2}(\tfrac{1}{\mu}L_{CS(\om)}+e(R-2\Lambda)).
\eeq
Thus in this case $\mu=\tfrac{g}{\kappa^2}$  and $v^2=\tfrac{16}{\kappa^2}$. The chiral point condition is $\mu l=1$ where $l$ is defined 
in terms of the cosmological constant as usual: $\Lambda=-\tfrac{1}{l^2}$. This implies that, to end up at a chiral point, the potential must  satisfy
\beq
V(v)=-\tfrac{1}{e}L_{X^6}(v)=-\tfrac{2\Lambda}{\kappa^2}=\tfrac{2}{\kappa^2 l^2}=\tfrac{2\mu^2}{\kappa^2}=\tfrac{2g^2}{\kappa^6}=
\tfrac{2g^2v^6}{16^3}=\tfrac{g^2v^6}{2\cdot 32\cdot 32},
\eeq
which differs from the background value above by a factor of 9. In \cite{Gran:2012mg} the observation was made that if two scalar fields are given the same VEV this factor of 9 disappears and one ends up at the chiral point with $\mu l=1$. In fact, by giving three scalar fields the same  VEV we find  instead that  $\mu l=3$ which has a null-warped solution.
Below we will elaborate on this situation and discuss the other values of $\mu l$ that appear.

The reason we expect the chiral point value $\mu l=1$, or other special values of $\mu l$, to play a role here is that we want to avoid  massive propagating gravity modes in the bulk \cite{Maloney:2009ck} which are not there in the conformal phase \cite{Gran:2008qx}. Introducing a similar kind of VEV in the ABJ(M) case \cite{Chu:2009gi}
leads, in fact, directly to the chiral point as 
%can be seen by inserting a real VEV for one of the scalar fields in the potential of the topologically gauged
%theory whose full potential 
was reviewed in the previous section. That special "critical" values of $\mu l$ are relevant for the broken phases also in  
${\mathcal N}=8$ theories will be a working hypothesis adopted in the following. This will  be crucial also for what kind of conformal field theories  that can arise at the boundary of the $AdS$ or the null Killing vector  backgrounds that we will find later.  Note that Minkowski  does also arise as a solution which
 may have a rather special "boundary CFT"  (see  \cite{Bagchi:2009my,Bagchi:2012yk} and references therein). We will not discuss boundary theories in any detail in this paper but we should mention here that one case that appears as a solution is the null-warped $AdS_3$ with its Schr\"odinger symmetries at the boundary discussed, e.g., in the context of cold atoms \cite{Son:2008ye,Balasubramanian:2008dm, Adams:2008wt}.

As just mentioned, one important aspect of the critical point of Li, Song and Strominger \cite{Li:2008dq} is that there are no massive gravity modes present.
The degeneration that occurs in the spectrum when tuning the non-critical TMG theory to its critical value may result in log-modes which would be problematic from a unitarity point of view\footnote{See, however, the previous footnote.} (see, e.g., the recent review \cite{Grumiller:2013at}). However, as explained in \cite{Maloney:2009ck} by choosing the boundary conditions one can consistently truncate the theory to a chiral subsector. A similar phenomenon may be at work  also in the null-warped case as argued in \cite{Anninos:2010pm}.
%Note that this issue is often discussed in the literature (see \cite{Maloney:2009ck,Grumiller:2013at} and references therein) without  involving any  scalar fields. 
The behavior of scalar
fields in this context has been discussed for instance in \cite{Bergshoeff:2012sc}. 
Other general  properties stemming from the fact that the theory comes from a   conformal phase may be extra symmetries as found for the null-warped metric (see below)\footnote{Extra symmetries  have, in fact,  also been found at the chiral point
\cite{Strominger:2008dp}.}.

 \subsection{Bosonic field equations and background solutions}

%%%%%%%%%%%%%%%%%%%%%%%%%%%%%%%%%%%%%%%%%%%%%%%%%%%%%%%
    
We here summarize the bosonic field equations found in \cite{Gran:2012mg}.  The Cotton equation reads
 \beqa
 &&\tfrac{1}{g}C_{\mu\nu}-\tfrac{eX^2}{16}(R_{\mu\nu}-\tfrac{1}{2}g_{\mu\nu}R)+
 \tfrac{e}{2}g_{\mu\nu}V(X)\notag\\[1mm]
 &&-\tfrac{e}{2}(D_{\mu}X^i_aD_{\nu}X^i_a-\tfrac{1}{2}g_{\mu\nu}D^{\si}X_a^iD_{\si}X_a^i)-\tfrac{e}{16}g_{\mu\nu}\Box X^2+\tfrac{e}{16}D_{\mu}D_{\nu}X^2=0.
 \eeqa
%  The equation of motion for the R-symmetry gauge fields, on the other hand, will be useful to have in more detail. Up to $\chi$ dependent terms it reads
% \beq
%\tfrac{1}{g} \ep^{\mu\nu\rho}G_{\nu\rho}^{ij}-eg^{\mu\nu}(\tilde D_{\nu}X_a^{[i})X_a^{j]}%+\tfrac{i}{8}e\bar\psi_a\ga^{\mu}\Ga^{ij}\psi_a
%=0.
% \eeq

 Turning to  the matter sector we first give the scalar field equation. Discarding the fermions  it becomes
 $\Box X^i_a- \tfrac{1}{8}X^i_a R-\partial_{X^i_a}\,V(X)=0$
 which can be  seen to be  consistent with the trace of  the Cotton equation.
%  In fact, combining these two scalar equations leads to the  condition on the potential
%$X\partial_XV(X)=6V(X)$ which is obviously correct in a three-dimensional  conformally invariant theory. 
Using the expression for  the potential  the  Klein-Gordon equation  becomes
\beqa
&& \Box X^i_a- \tfrac{1}{8}X^i_a R=%-\tfrac{\lambda^2}{2}\ep^{abcg}X_b^jX_c^kf^{defg}X^i_dX^j_eX^k_f
\notag\\[1mm]
&&\tfrac{g^2}{32 \cdot 32}(3X^i_a(X^2)^2-8X_a^i(X_b^jX_b^k)(X_c^jX_c^k)-16X^2X_a^kX_b^kX_b^i+48X_a^j(X_b^jX_b^k)(X_c^kX_c^i)).\notag\\[1mm]
%&&-\tfrac{i\lambda}{2}\bar\psi_c\Ga^{ij}\psi_dX^j_b\ep^{abcd}+\tfrac{ig}{32}(\bar\psi_b\psi_bX^i_a-10\bar\psi_a\psi_bX^i_b+2\bar\psi_a\Ga^{ij}\psi_bX^j_b)
\eeqa
%The field equation for the gauge field $A_{\mu}^{ab}$  is, again discarding the fermion terms,
%\beq
%\tfrac{2}{g} \ep^{\mu\nu\rho}F_{\nu\rho}^{ab}+e\,g^{\mu\nu}(\tilde D^{\mu}X^i_c)\de^{ab}_{cd}X^i_d%+\tfrac{i}{2}e\bar\psi_c\ga^{\mu}\psi_d\tilde M_{cd}^{ab}
%=0,
%\eeq
Finally, for the R-symmetry gauge field we have  the following field equation

 \beq
 \ep^{\mu\nu\rho}\,G_{\nu\rho}^{ij}+ g\,e g^{\mu\nu} X^{[i}_aD_{\nu}X^{j]}_a=0,
 \eeq
while for the $SO(N)$ gauge field $A_{\mu}^{ab}$ we get
 \beq
-2 \ep^{\mu\nu\rho}\,F_{\nu\rho ab}+ g\,e g^{\mu\nu} X^{i}_{[a}D_{\nu}X^{i}_{b]}=0.
 \eeq
 The  field equations for the two vector  fields are trivially satisfied in the backgrounds we  use here. Thus we can concentrate our efforts on the Cotton and Klein-Gordon equations.

We now demonstrate that these last equations allow for a number of different  background solutions
%\footnote{Solutions of TMG in general were discussed in \cite{Maloney:2009ck}. }
two of which were briefly mentioned but not
analysed in \cite{Gran:2012mg}. The first step will be to solve the Klein-Gordon equation.
To do this we introduce a VEV $p\times p$ unit matrix $v{\bf 1_{p\times p}}$ by setting\footnote{There may be other ways to introduce scalar VEVs. Only some simple modifications 
of the VEV used here have been checked and seen to give nothing new.} 
\beq
X^i{}_a=<X^i{}_a>+x^i{}_a=v\de^I_A+x^i{}_a,
\eeq
where the VEV term proportional to $\de^I_A$ ($I=1,2,..,p,\,\,A=1,2,...,p\leq 8$ or $p\leq N$ if $N<8$) means that the scalar fields that are given the same  VEV $v$ are the first $p$ ones along the diagonal starting from the upper left-hand corner of the rectangular matrix $X^i{}_a$ having 8 rows and $N$ columns. Recall that the indices take the values
$i=1,2,...,8$ and $a=1,2,...,N$ where $N$ can be any positive integer. The capital indices $A, B,..$ and $I, J,..$ are thus of the same kind as far as their transformation properties are concerned and we will not distinguish between them  from now on. $x^i{}_a$ are the fluctuations relative these VEVs. We  thus  have, e.g.,
$
X^2=X^i{}_aX^i{}_a
%=pv^2+v\de^I_Ax^I_A+x^2
=pv^2+2vz+x^2,
$
where  the trace $x^I{}_I=z$ and $x^2=x^i{}_ax^i{}_a$.

For the index choice $i=I,a=A$ the scalar field equation in the background of the matrix VEV becomes 
%\beq
%-\tfrac{1}{8}v\delta^I_A\,\bar R=\tfrac{g^2}{32 \cdot 32}(3vp^2\delta^I_Av^4-8vp\delta^I_Av^4-16v^2p\delta^I_Av^3+48v^5\delta^I_A),
%\eeq
%When simplified this becomes
\beq
\bar R=6\Lambda=-\tfrac{6}{l^2}=-\tfrac{6}{16 \cdot 16}g^2v^4(p-4)^2,
\eeq
where $\bar R$ refers to the background value of the curvature scalar. 
This equation will be a constraint valid in all considerations to be made in the rest of the paper whether the background is maximally symmetric or not.
%which when used in the lagrangian at order $v^5x$ above gives
%\beq
%L|_{\mathcal O(x^1)}=-\tfrac{e}{16}\tfrac{2v}{\sqrt{p}}z(-\tfrac{6}{16 \cdot 16}g^2v^4(1-\tfrac{8}{p}+\tfrac{16}{p^2}))-\tfrac{6\,e\,g^2}{2\cdot 32\cdot 32}\tfrac{v^5}{\sqrt{p}}z(1-\tfrac{8}{p}+\tfrac{16}{p^2})=0
%\eeq
%i.e., it vanishes for any value of $p$ as expected.
In order to discuss the other scalar equations we split the  indices as follows:
\beq
i=(I,\hat i),\,\,\,\,a=(A,\hat a).
\eeq
 We then note  that using $i=I,a=\hat a$ etc, the remaining scalar field equations are trivially satisfied since there are no VEVs connecting the two indices in these cases. 

What remains to be solved is the Cotton equation. To do this for general values of $p$ we consider first the Lagrangian with the background put in for all fields except the metric:%, or in other words
%we need to construct the potential at order $v^6$.
%By expanding the lagrangian in powers of the VEV $v$ we can check the gravitational field equations order by order in the VEV. In the rest of this section we consider the potential at order $v^6$ and 
%$v^5$ of which the latter is equivalent to  the Cotton equation. The mass spectrum requires an analysis of the $v^4$ terms in the potential which is the main subject of the next section.
%At order $v^6$ the lagrangian reads
%\beq
%L_{v^6}=-\tfrac{1}{g}L_{CS(\omega)}+\tfrac{2}{g}L_{CS(B)}-\tfrac{4}{g}L_{CS(A)}-\frac{ev^2}{16}\,R-\tfrac{ev^6g^2}{2\cdot 32\cdot 32}(1-\tfrac{4}{p})^2
%\eeq
%To lowest order in all fields except the metric we have
\beq
L_{VEV}=-\tfrac{1}{g}L_{CS(\omega)}-\frac{epv^2}{16}\,R-\tfrac{ev^6g^2}{2\cdot 32\cdot 32}p(p-4)^2.
\eeq
Comparing this to the Lagrangian used in the  analysis of LSS \cite{Li:2008dq}
%\footnote{In this reference one uses the gravitational Chern-Simons term
%written in terms of the affine connection. However, for our purposes this is equivalent to using a Chern-Simons term constructed from the spin connection. For a discussion 
%on the relation between these two forms of the gravitational Chern-Simons term, see e.g., \cite{Guralnik:2003we}. }
\beq
L=-\tfrac{1}{\kappa^2}(\tfrac{1}{\mu}L_{CS(\omega)}+e\,(R+\tfrac{2}{l^2})),
\eeq
we can read off its parameters  expressed in terms of our variables $v,g$:
\beq
\kappa^2\mu=g,\,\,\,\kappa^2=\tfrac{16}{pv^2},\,\,\,\tfrac{2}{\kappa^2l^2}=p(p-4)^2\tfrac{v^6g^2}{2\cdot 32 \cdot 32 },
\eeq
where $\kappa^2$ and $l$ have dimension $L^1$ and $\mu$  dimension $L^{-1}$ since $g$ is dimensionless. Recall that the field $X^i{}_a$ and thus $v$ has dimension $L^{-1/2}$.
The parameter  relations above can be written
\beq
\mu=\tfrac{g}{\kappa^2}=\tfrac{gpv^2}{16},\,\,\,l=\tfrac{1}{\kappa}\tfrac{2\cdot32}{|p-4|\sqrt{p}v^3g}=\tfrac{16}{|p-4|gv^2},
\eeq
and hence
\beq
\mu l=|1-\tfrac{4}{p}|^{-1}.
\eeq
This equation gives the following values  for $p=1,2,...,8$: 
\beq
\label{mu-l-values}
\mu l=\tfrac{1}{3}, 1, 3, \infty, 5, 3, \tfrac{7}{3}, 2.
\eeq

The interesting cases are $p=2$ which allows for an ordinary critical (chiral) round $AdS$ solution together with $p=3$ and $p=6$ both having a null-warped $AdS$  (see \cite{Anninos:2008fx,Anninos:2010pm} and references therein)
   as a possible solution. This latter solution has a non-zero Cotton tensor but
a constant curvature scalar as we saw above is a  property  all  solutions  must satisfy.
%\footnote{Null-warped manifolds also appear as solutions in theories
%without a gravitational Chern-Simons term. In, for instance, \cite{Detournay:2012dz} they are made possible due to the presence of massive vector fields. These models can also be embedded  in string theory.}.
 Also $p=4$ is 
interesting since  the potential vanishes and the solution is flat Minkowski space-time. Recent work like \cite{Bagchi:2009my,Bagchi:2012yk} might be relevant in this case. These geometries are all very well-known and will be described briefly below. However, 
 for $p=5$  we get $\mu l=5$ which is  intriguing: a solution with $\mu l=5$ was discovered only   recently by Ertl, Grumiller and Johansson (EGJ) \cite{Ertl:2010dh} and as we will see below the way  this solution is obtained is very different from the other ones mentioned here. 
 %However, using the relation between the existence of supersymmetry and a null Killing vector \cite{Gibbons:2008vi} it should be compatible with supersymmetry for %one of the possible orientations (or sign of $\ep^{012}$) of the the space-time manifold.
% \footnote{Note that the sign in front of the whole lagrangian quoted here is opposite to that used in \cite{Li:2008dq}. This may have important consequences for the unitarity in non-supersymmetric backgrounds  since black holes will have negative energy with the sign used in our case. This sign cannot be 
%changed since it is related to the sign of the Klein-Gordon term in the lagrangian and we assume that the original ungauged matter theory is well-behaved, in particular unitary.}.

 Thus several of the $\mu l$ values  in the list above can be connected to
 solutions of TMG that are critical or in some sense special, at least this is the case for $p=2,3,4,5,6$. It is therefore natural to wonder  if the remaining values also have special solutions which, however, have not yet been found in TMG\footnote{The value $\mu l=2$ does in fact come up in the context of $BTZ$ black holes \cite{Afshar:2010ii}. I am grateful to H.R. Afshar for pointing this out to me. See also {\bf Note added} at the end of the last section.}. Note that non-critical solutions based on the round $AdS_3$ exist for any value of $\mu l$ but then there are propagating massive (positive energy) gravitons. 
 In this context we may remind the reader that the theories discussed here have a potential problem with unitarity due to negative energy black holes and boundary modes. For a discussion of this issue in bosonic TMG, see \cite{Deser:2010df}. Some  perhaps relevant comments concerning  supersymmetric  theories can be found in \cite{Becker:2009mk}.

\subsection{Some properties of the special (critical) solutions}

%%%%%%%%%%%%%%%%%%%%%%%%%%%%%%%%%%%%%%%%%%%%%%%%%%%%%%%

In this subsection we discuss some of the special solutions of the Cotton equation that are possible for the values of $\mu l$ that appeared for the different choices of scalar VEVs.
%While the critical round $AdS_3$ ($\mu l=1$) and the null-warped $AdS_3$ ($\mu l=3$) have been studied in great detail in the past, the metric for $\mu l=5$
%was not discovered until recently by Ertl et al \cite{Ertl:2010dh} and is not yet known in closed form. 
There are several recent attempts to classify the known solutions of TMG,
see for instance \cite{Gibbons:2008vi,Chow:2009km, Chow:2009vt} and \cite{Ertl:2010dh}\footnote{A complete classification of all homogenous solutions with constant scalar invariants in TMG, NMG and GMG \cite{Bergshoeff:2009hq} can be found in \cite{Siampos:2013foa}.}. These papers  also contain some new solutions as well as most of the original references for the previously found   solutions which  appear  in various guises in the literature. 
E.g., in  \cite{Chow:2009km} the Petrov-Segre classification is adapted to this situation and shown to directly  account for the known solutions as belonging to a very limited set of classes.  We will, however,  be mostly concerned with  a  method discussed first by Clement \cite{Clement:1994sb} and later used in  \cite{Ertl:2010dh}.  In the latter  work  the authors divide the construction of stationary axi-symmetric TMG solutions into 
 sectors called Einstein, Schr\"odinger, warped and generic. After observing that all known solutions belong to the first three classes they go on to construct a new solution that belongs to the general class and which turns out to have rather special properties. The metric has $\mu l=5$ and is non-polynomial in the radial coordinate $r$ (see below). % (see, however, below for a possible connection to solutions of 
%NMG (New Massive Gravity) \cite{Bergshoeff:2009hq})

%We will below make heavy use of the paper by Ertl et al \cite{Ertl:2010dh}. It is thus appropriate to mention the fact that the solutions that we in this paper focus on and call critical
%are the ones with $\mu l=1,3,5,\infty$, and these are precisely the ones that Ertl et al find to be somewhat special for reasons that seem to have nothing to do with the fact that they have a common origin as solutions
%of the topologically gauged free $CFT_3$ theory discussed here. This connection to a superconformal theory is one of the main  results in this note.

It is convenient to use light-cone coordinates such that three-dimensional Minkowski space-time with signature ($-++$) is described by the metric
\beq
ds^2=d\rho^2+2dudv,
\eeq
where $\rho$ is the "radial" coordinate taking values from $-\infty$ to $+\infty$ and $2dudv=-dt^2+dx^2$. 

In the literature
other  closely related  coordinates appear: for instance the coordinate $r$ ($0<r<\infty$) related to $\rho$  by
 $2\rho/l=log(r/l)$ is often used.  Note, however, that in the reference  \cite{Ertl:2010dh} $\rho$ corresponds to our
radial  coordinate $r$. Also, the commonly used coordinate $z$ can then be introduced by $r/l=z^{-2}$.
%\footnote{The perhaps most common notation for this coordinate is $z$.}

The existence of global coordinate systems that turn the Poincar\'e patch into a geodecically complete space are very important in the cases below. This is one of the features that may be common to all the solutions that we call "critical" in this paper. The global coordinates for  the round $AdS_3$ are well-known and  the null-warped case is thoroughly discussed in \cite{Blau:2009gd} while the situation for the EGJ solution with $\mu l=5$ is not clear.\\

%%%%%%%%%%%%%%%%%%%%%%%%%%%%%%%%%%%%%%%%%%%%%%%%%%%%%%%

{\bf Critical $AdS_3$ ($p=2$, $\mu l=1$)}:
%%%%%%%%%%%%%%%%%%%%%%%%%%%%%%%%%%%%%%%%%%%%%%%%%%%%%%%
The  metric for the round $AdS_3$ with radius $l$ is
\beq
ds^2=d\rho^2+2e^{2\rho/l}dudv=l^2\frac{dr^2}{4r^2}+\frac{2r}{l}dudv=\frac{1}{z^2}(l^2dz^2+2dudv).
\eeq
%where we also used $\tfrac{r}{l}=z^{-2}$. 
Criticality refers in this case  to the fact that the massive bulk gravity mode disappears and a potentially chiral  boundary theory becomes possible as $\mu l$ is tuned to one \cite{Li:2008dq, Maloney:2009ck}. In the context of this paper with a large number of scalar fields present, the chiral limit should be reconsidered. Some relevant results in this direction may be found in \cite{Bergshoeff:2012sc}. Since in three dimensions the Weyl tensor vanishes, the Riemann tensor is given entirely by the traceless Ricci tensor
and the curvature scalar. It then follows that being  Einstein  is equivalent to being maximally symmetric, and hence
the above metric is the unique solution of TMG with zero Cotton tensor\footnote{For non-Einstein solutions with $\mu l=1$, see
\cite{Compere:2010xu}. See also \cite{Chow:2009km}.}. This corresponds to the class ${\mathcal O}$
in the Petrov-Segre classification in \cite{Chow:2009km} and to the Einstein sector 
in \cite{Ertl:2010dh}. 
%The round $AdS_3$ solves the TMG equations of motion for any value of $\mu l$ and the criticality issue only comes in when discussing
%the spectrum in the bulk and the type of $CFT_2$ that appears on the boundary. The property of interest for us in this note is the possibility to get rid of all propagating gravity modes in the bulk by choosing the boundary conditions appropriately, see for instance \cite{Maloney:2009ck} for a thorough discussion of this
%subtle point. This we believe is an effect of being related to a conformal phase where there  are no such modes.

In \cite{Gibbons:2008vi} the Killing spinor equation is solved and shown to have two solutions corresponding to the two components of a
spinor in three dimensions. Thus this background allows for eight ordinary $AdS_3$ supersymmetries in the context of this paper.\\
%The explicit transformation rules will be discussed in the next section.

%%%%%%%%%%%%%%%%%%%%%%%%%%%%%%%%%%%%%%%%%%%%%%%%%%%%%%%

{\bf Null-warped $AdS_3$ or $Sch_3(z=2)$ ($p=3$ and $ 6$, $\mu l=3$)}:
%%%%%%%%%%%%%%%%%%%%%%%%%%%%%%%%%%%%%%%%%%%%%%%%%%%%%%%
The relation $z=\tfrac{\mu l +1}{2}$ is obtained by using  the ansatz $ds^2=d\rho^2+2e^{2\rho/l}dudv\pm e^{2z\rho/l}du^2$ to solve the Cotton equation in TMG.  
For  the value $2$ of the dynamical scaling parameter $z$, corresponding to $\mu l=3$,  the solution is critical
%\footnote{The issue of log modes arises only for $|\mu l|=1$.} 
 in the sense that among the solutions with a null Killing vector it has no tidal forces, a global coordinate system \cite{Blau:2009gd} and an extra conformal generator \cite{Son:2008ye,Balasubramanian:2008dm}\footnote{This fact is important for the  condensed matter applications to  Fermi gas/cold atom systems. For other  properties of this geometry relevant for applications, see \cite{Adams:2008wt, Guica:2010sw, Kraus:2011pf,Wang:2013tv}.}. As discussed in \cite{Anninos:2010pm}, it may also be possible to truncate the 
 spectrum in a chiral fashion similar to the $\mu l=1$ case of the previous subsection.
%there is a compact direction involved in this geometry and which corresponds to particle number (see 1007.2184 and
%Mahajana's thesis, p 10-11). This fact is crucial for many of the  applications  of this particular $z=2$  Schr\"odinger geometry, see, e.g., \cite{Son:2008ye, Balasubramanian:2008dm, Adams:2008wt, Guica:2010sw, Wang:2013tv}. } compared to the other cases in the set of geometries $Sch_z$, the Schr\"odinger geometries 
This metric can be written as follows
\beq
ds^2=d\rho^2+2e^{2\rho/l}dudv\pm e^{4\rho/l}du^2=l^2\frac{dr^2}{4r^2}+\frac{2r}{l}dudv\pm \frac{r^2}{l^2}du^2=\frac{1}{z^2}(l^2dz^2+2dudv)\pm\frac{du^2}{z^4},
\eeq
where the properties of this geometry depend on the sign in front of the last term, see \cite{Anninos:2008fx}. 
%In fact, using the minus sign gives a well-behaved metric \cite{Anninos:2008fx} and then by identifying $u$ with time $t$  one obtains the solution that is of interest in condensed matter applications 
%\cite{Son:2008ye,Balasubramanian:2008dm,Wang:2013tv}.
%Note that this metric belongs to the class called Schr\"odinger also in \cite{Ertl:2010dh}. The more common name "null-warped" is in fact somewhat unfortunate: 
%it is, e.g.,  not a limit of the warped metrics with $\mu l\neq 3$ as explained in \cite{Anninos:2008fx}\footnote{Taking this limit gives in fact back the round $AdS_3$\cite{Anninos:2008fx}.  The warped metrics can be either squashed or stretched in either the time direction or a space direction. The named "warped" should be replaced by "squashed" or "stretched" 
%since there are no functions involved
% as remarked in \cite{Chow:2009km}. In \cite{Ertl:2010dh} (see the next case $\mu l=5$) the null-warped case does indeed arise as a limiting case among the warped ones but there the parameter that is tuned is again $z$ not the amount of squashing.}.

In this case we know from \cite{Gibbons:2008vi} that the three-dimensional geometry can only support one (component) supersymmetry due to the presence of
a null Killing vector (without being the round $AdS$). In fact, the existence of a Killing spinor in this geometry implies that there is a null  vector $K_{\mu}$ satisfying
\beq
D_{\mu}K_{\nu}=-\ep_{\mu\nu\rho}K^{\rho}.
\eeq
Turning the argument around \cite{Gibbons:2008vi}, assuming a null Killing vector (not necessarily satisfying the anti-symmetric part of the above equation), the  TMG geometries are just 
 the supersymmetric ones given above  allowing, however, also for  their orientation flipped versions. \\

%It is interesting to note that there is a compact direction involved in this geometry and which corresponds to particle number (see 1007.2184 and
%Mahajana's thesis, p 10-11). This fact is crucial for many of the  applications  of this particular $z=2$  Schr\"odinger geometry, see, e.g., \cite{Son:2008ye, Balasubramanian:2008dm, Adams:2008wt, Guica:2010sw, Wang:2013tv}.
%conformal group see 0706.3746.

%There is something called the Schr\"odinger soliton in App A of \cite{Adams} 1103.3472. Is this the analogue of AdS, ie
%the usual Schr\"odinger background ???

%One should note that the round AdS is also a solution for the value $\mu l=3$ as for any other value. However, for other values than $\mu l=1$ there is 
%a massive positive energy graviton in the spectrum. Had the over-all sign of the lagrangian been opposite  to the one used in this paper (which is dictated by the
%sign of the Klein-Gordon term and that the scalars are physical and unitary) this massive graviton would have had negative energy and given rise to a 
%break-down of space-time  as discussed in \cite{Lashkari:2010iy}. Here, on the other hand, that scenario is not relevant and we will, as
%mentioned previously, just assume that since the conformal phase has no bulk gravitons this will continue to be true, or at least exclude phases which do have massive gravitons.
%If the relation to a conformal phase has any new information to offer concerning the question of boundary conditions and log-modes remains an interesting question.

%%%%%%%%%%%%%%%%%%%%%%%%%%%%%%%%%%%%%%%%%%%%%%%%%%%%%%%

{\bf The EGJ solution ($p=5$, $\mu l=5$ )}:
%%%%%%%%%%%%%%%%%%%%%%%%%%%%%%%%%%%%%%%%%%%%%%%%%%%%%%%
This solution was first obtained by Ertl, Grumiller and Johansson (EGJ) in \cite{Ertl:2010dh}\footnote{For an earlier analysis using these methods, see \cite{Maloney:2009ck}.} using an approach discussed originally by Clement
\cite{Clement:1994sb}.  To find all solutions of TMG that are stationary and axi-symmetric one may  adopt the following ansatz for the metric:
\beq
ds^2=(det\, h)^{-1} d r^2+h_{\al\be}dx^{\al}dx^{\be}=(det \,h)^{-1} d r^2+h_{++}dudu+2h_{+-}dudv+h_{--}dvdv,
\eeq
where the three functions in the  $h_{\al\be}$ part of the metric depend only on the radial coordinate $r$. (In this subsection we use  the conventions   of \cite{Ertl:2010dh} apart from renaming their coordinate $\rho$ as $r$ and denoting derivatives  by a prime instead of an over-dot.)
Thus we denote the functions $h_{++}, h_{--}$ and $h_{+-}$ as $X^+,X^-,Y$, respectively,
and note that
\beq
det \,h = X^+X^--Y^2:=X^iX^j\eta_{ij},
\eeq
defines an auxiliary flat metric $\eta$ with signature $(+,-,-)$. Setting  $X^i=(X^+,X^-,Y)$,
we find  that (the physical) Minkowski space corresponds to $X^i=(0,0,1)$ and the maximally symmetric $AdS_3$ to $X^i=(0,0,r)$ while the null-warped case is obtained from $X^i=(r^2,0,r)$.   
In all these cases ${\bf X^{'}}\cdot {\bf X^{''}}=0$ and ${\bf X^{''}}^2=0$ which can be shown to imply ${\bf X^{'''}}=0$. As emphasized in  \cite{Ertl:2010dh} the first  two  conditions reduce the phase space to
a four-dimensional hypersurface. The new solution with $\mu l=5$ will not satisfy these conditions and therefore seems to make use of the entire six-dimensional phase space. The functions $X^i$ then no longer satisfy ${\bf X^{'''}}=0$ and will, in fact, become non-polynomial in the radial coordinate. No closed form of the solution is yet known. 
%Nevertheless, since this solution has a null Killing vector it should be compatible with half supersymmetry (for one of the possible orientations) according to the argument given above for $\mu l=3$ 
%\cite{Gibbons:2008vi}.

The set of equations  obtained by using this ansatz for the metric in the Cotton equation divides into a hamiltonian constraint, which involves fields acted upon by at  most two  $r$ 
derivatives (see below), and three  equations for the $X^i$ containing terms that are third order in derivatives. However, one can integrate the third order equations once by employing the fact that  the
"angular momentum" associated to the Lorentzian symmetry of the dynamical equations containing $\eta_{ij}$ and ${\bf X}$ is a constant of motion. In fact, acting with a derivative on
\beq
{\bf J}={\bf X}\times{\bf { X^{'}}}+\tfrac{1}{\mu}{\bf X}\times({\bf X}\times{\bf X^{''}})-\tfrac{1}{2\mu}{\bf{X^{'}}}\times({\bf { X}}\times{\bf {X^{'}}}),
\eeq
 results in a cross product of ${\bf X}$ and the third order equations of motion. Thus one wants to solve this last equation together with the following  second order equation, which is the hamiltonian constraint in the TMG theory,
\beq
\tfrac{1}{2}{\bf X^{'}}^2+\frac{2}{l^2}-\frac{1}{\mu}\ep_{ijk}X^iX^{'j}X^{''k}=0.
\eeq
In fact, all the dynamical  equations follow from the following TTM (topologically massive mechanics) action \cite{Clement:1994sb}:
\beq
 S_{TMM}=\int d \rho\,e(\tfrac{1}{2}e^{-2}{\bf X^{'}}^2-\frac{2}{l^2}-\frac{1}{2\mu}e^{-3}\ep_{ijk}X^i X^{'j} X^{''k}),
\eeq
where also an einbein $e$ has been introduced.

 These equations also imply that that the curvature scalar
can be expressed as
\beq
R=2{\bf X}\cdot{\bf X^{''}}+\frac{3}{2}{\bf X^{'}}^2=-\frac{6}{l^2},
\eeq
which, if combined with the hamiltonian constraint, implies
\beq
\mu {\bf X}\cdot{\bf X^{''}}+{\bf X}\cdot{\bf X^{'}}\times{\bf X^{''}}=0.
\eeq

Following Ertl et al. \cite{Ertl:2010dh} for $\mu l=5$, if  we set ($s=0,\pm$1)
\beq
{\bf X^T}|_0=(1,0,0),\,\,{\bf X^{'T}}|_0=\mu(s,0,\tfrac{2}{5}),
\eeq
we can start solving the equations in an iterative fashion. We find
\beq
{\bf X^{''T}}|_0=({X}^{''+}|_0,0,Y^{''}|_0),
\eeq
where 
%\beq
%Y^{''}|_0=\tfrac{1}{2\mu}X^{'''-}|_0,\,\,\,X^{''+}|_0=\tfrac{25}{2\mu}(\tfrac{29\mu s}{20}+\tfrac{5}{4\mu}Y^{''}|_0)Y^{''}|_0-\tfrac{125s}{16\mu}X^{'''-}|_0,
%\eeq
%which simplifies to
\beq
Y^{''}|_0=\tfrac{1}{2\mu}X^{'''-}|_0,\,\,\,X^{''+}|_0=\tfrac{125}{32\mu^4}(X^{'''-}|_0)^2+\tfrac{5s}{4\mu}X^{'''-}|_0.
\eeq
Thus, one difference between this solution and  the  critical ones discussed above is that the  component $X^-$ is non-zero starting at third order in $r$.
How this affects the possibility for this geometry to support supersymmetry remains to be clarified.\\
%A non-zero $X^-$ is probably not  an artifact of the method of construction and thus the argument that metrics allowing for a null Killing vector  supports supersymmetry 
% for one of the orientations  \cite{Gibbons:2008vi} can  not be used.

%This method of solving the TMG equations 
%for higher values of $\mu l$ will be 
%briefly discussed in the concluding section.

%%%%%%%%%%%%%%%%%%%%%%%%%%%%%%%%%%%%%%%%%%%%%%%%%%%%%%%

{\bf Minkowski ($p=4$, $\mu l=\infty$}):
%%%%%%%%%%%%%%%%%%%%%%%%%%%%%%%%%%%%%%%%%%%%%%%%%%%%%%%
Recall that we are in this paper assuming that the relevant solutions are in some sense  "critical" with properties that stem for their connection to a conformal phase.
In the context of Minkowski space this is a particularlry delicate issue. However, we note that there are discussions
in the literature concerning the  
 possibility to tune an $AdS$ bulk geometry to a flat space and follow what happens to the symmetries of the CFT at the boundary,
 see, e.g., \cite{Bagchi:2009my,Bagchi:2012yk}.
 %(1208.1658). 
   This could  be telling us to define a "critical" Minkowski solution  for $p=4$ by relating it to the BMS algebra, see, e.g., \cite{Barnich:2012aw}.\footnote{In relation to the second of these references we note 
   that the limit  used there to  get the wanted flat space CFT is similar to tuning the VEV $v$ introduced here  to zero keeping $g$ fixed!}

\section{Mode analysis and supersymmetry}

%%%%%%%%%%%%%%%%%%%%%%%%%%%%%%%%%%%%%%%%%%%%%%%%%%%%%%%

%The purpose of this section is discuss some aspects of the spectrum of physical states including how they get organized  after the breaking of the 
%conformal symmetries. This will be done for a general value of the parameter $p$ which we  introduced in the previous section as the number of scalar fields
%getting the VEV $v$. There we also saw how $p$ is related to $\mu l$  which  specifies what kind of  critical vacuum is obtained. 

%%%%%%%%%%%%%%%%%%%%%%%%%%%%%%%%%%%%%%%%%%%%%%%
   
%\subsection{Symmetry breaking: The Chern-Simons sector}

%%%%%%%%%%%%%%%%%%%%%%%%%%%%%%%%%%%%%%%%%%%%%%%

To study the spectrum we should  expand the Lagrangian and the field equations around the VEV $v$ using
\beq
X^i{}_a=<X^i{}_a>+x^i{}_a=v\delta^I_A+x^i{}_a,
\eeq
where the  VEV matrix is proportional to the $p\times p$ unit matrix, i.e., $A, I=1,2,...,p\leq 8$ (or  $p\leq N$ if $N<8$).
Note that  we have defined the upper index as the first one and the lower as the second one (whether indices are upper or lower will not matter from now on) and
 that we in the broken phases do not need to distinguish between the two sets of capital  Latin  indices $A,B,...$ and $I,J,..$.
As already mentioned we define also  the remaining index values $\hat i$ and $\hat a$  by setting 
$
i=(I,\hat i),\,\,a=(A,\hat a).
$
%where, as just mentioned, capital indices are in the VEV directions. 
\subsection{Symmetry breaking and massive vector fields}

At this point we can insert the  VEV into the Klein-Gordon term in the Lagrangian to  determine the symmetry breaking pattern. 
The terms proportional to $v^2$ are 
\beq
\label{Lv-square}
L(v^2)=-\thalf v^2(A_{\mu}^{ ab}\de^I_B+B_{\mu}^{ij}\de_A^J)^2=-\thalf v^2((A_{\mu }^{\hat a B})^2+(B_{\mu}^{ \hat i J})^2+(A_{\mu}^{ AB}-B_{\mu }^{AB})^2),
\eeq
where a square  $(A_{\mu}{}^{ AB})^2=A^{\mu AB}A_{\mu}^{AB}$ etc. Note also that we have adopted the summation 
rule that $A_{\mu}^{ ab}\de^I_B:=A_{\mu}^{ aB}\de^I_B$ etc. Thus
the symmetry breaking of the bosonic gauge and R-symmetries is governed by the coset
\beq
G/H:\,\,\,G=SO(N)\times SO(8),\,\,\,H=SO(N-p)\times SO(8-p)\times SO(p)_{diag},
\eeq
where the factor $SO(p)_{diag}$ is the diagonal part of the two $SO(p)$ groups coming from $SO(N)$ and $SO(8)$ after breaking. 

However, the two gauge fields
involved in the $SO(p)$ part of this system have  differently normalized CS terms and the equations of motion need to be properly diagonalized to find the actual mass of the higgsed vector field.
The combination of the two vector fields that remains a gauge field after breaking is determined as follows. The linearized vector equations read, for the R-symmetry gauge field, with $\de B_{\mu}:=b_{\mu}$ and $\de A_{\mu}:=a_{\mu}$,
\beq
2\bar\ep_{\mu}{}^{\nu\rho}\pd_{\nu}b_{\rho}+gv^2\bar e \de^{\rho}_{\mu}(a_{\rho}-b_{\rho})=0,
\eeq
and for  the $SO(p)$ gauge field
\beq
4\bar\ep_{\mu}{}^{\nu\rho}\pd_{\nu}a_{\rho}+gv^2\bar e \de^{\rho}_{\mu}(a_{\rho}-b_{\rho})=0.
\eeq
As we will now see, the reduction of this system to a single vector field is similar  to that used by  Mukhi and Papageorgakis in \cite{Mukhi:2008ux} but will here in addition to the Yang-Mills term  generate  a topological mass term  in a curved background.  %This is a consequence of the fact that the mass term is not a priori
%orthogonal to the massless gauge field. 
To see this  we define 
%rewrite the above equations in terms of the two orthogonal vector fields 
\beq
c'_{\mu}=2a_{\mu}-b_{\mu},\,\,\,\,c_{\mu}=a_{\mu}+2b_{\mu},
\eeq
which satisfy %{\it CHECK: background covariance!!!}
\beq
\bar \ep_{\mu}{}^{\nu\rho}\pd_{\nu}c'_{\rho}=0,\,\,\,\,\bar \ep_{\mu}{}^{\nu\rho}\pd_{\nu}c_{\rho}=-\tfrac{5m}{4}\bar e(a_{\mu}-b_{\mu})=-\tfrac{m}{4}\bar e(3c'_{\mu}-c_{\mu}),
\eeq
where the mass $m=gv^2$. In the parity symmetric case studied in \cite{Mukhi:2008ux} the field $c_{\mu}$ does not appear on the right hand side of the second equation. The general non-symmetric situation with arbitrary parameters in front of the various terms is, however,  discussed in \cite{Mukhi:2011jp} and  contains the features seen here. Eliminating  the field $c'_{\mu}$ we obtain in our case the following field equation for $\bar H_{\mu\nu}^{IJ}=\pd_{\mu}c_{\nu}-\pd_{\nu}c_{\mu}$:
\beq
\bar e\,\bar D^{\nu}\bar H_{\nu\mu}=\frac{m}{8}\bar \ep_{\mu}{}^{\nu\rho}\bar H_{\nu\rho},
\eeq
which is  a topologically massive gauge theory \cite{Deser:1982vy,Townsend:1983xs} in a curved background.\footnote{For a very nice discussion of the various mass terms that appear in this context and the relations between them, see \cite{Mukhi:2011jp}.}  Thus the Yang-Mills coupling constant 
$g^2_{YM}$ is proportional to the mass parameter $m=gv^2$.
%This is similar to the higgs effect introduced in \cite{Mukhi:2008ux}
%which, however, produces a massless Yang-Mills theory  since the starting point there is  the parity symmetric CS sector of (flat space) BLG theory.

The last task concerning the vector fields is to rewrite the covariant derivative in terms of the gauge field $c_{\mu}$ which will also give us a hint about  the structure of the full non-abelian case. Thus, using the above expression for $c'_{\mu}$, we get
\beq
a_{\mu}=\frac{1}{5}(c_{\mu}+2c'_{\mu})=\frac{1}{3}(c_{\mu}-\frac{8}{5m\bar e}\bar \ep_{\mu}{}^{\nu\rho}\bar D_{\nu}c_{\rho}),
\eeq
and
\beq
b_{\mu}=\frac{1}{5}(c_{\mu}+2c'_{\mu})=\frac{1}{3}(c_{\mu}+\frac{4}{5m\bar e}\bar \ep_{\mu}{}^{\nu\rho}\bar D_{\nu}c_{\rho}).
\eeq
Now we rescale $c_{\mu}$ to cancel the factor $\frac{1}{3}$, rename the field as  $C_{\mu}$ and express the  covariant derivative  as follows
\beqa
&&D_{\mu}X^I{}_A=\pd_{\mu} X^I{}_A+A_{\mu AB}X^I{}_B+B_{\mu}^{IJ}X^J{}_A\rightarrow \cr
&&D_{\mu}X^I{}_A-\frac{4}{5me}(2(\ep D C^{IJ})_{\mu}X^J{}_A
-(\ep D C_{AB})_{\mu}X^I{}_B),
\eeqa
where the new covariant derivative, also denoted $D_{\mu}$, is
\beq
D_{\mu}X^I{}_A=\pd_{\mu} X^I{}_A+C_{\mu AB}X^I{}_B+C_{\mu}^{IJ}X^J{}_A.
\eeq
A  more complete treatment using the  non-abelian field strength $H_{\mu\nu}^{IJ}$ defined by the commutator as usual is obtained by the replacement
\beq
(\ep D C^{IJ})_{\mu}\rightarrow \thalf\ep_{\mu}{}^{\nu\rho}H_{\nu\rho}^{IJ}.
\eeq
%and similarly for other expressions of this type.
%\footnote{With the "background" type  expansion $A=C+\de A$ and $B=C+\de B$, 
%the CS term for $A$ becomes
%\beq
%L_{CS(A)}=L_{CS(C)}+\de AH(C)+\thalf \de AD_C\de A+\frac{1}{3}\de A^3,
%\eeq
%and similarly for $B=C+\de B$
%. Then setting $\de A_{\mu}= -2\de B_{\mu}= -\frac{4}{5me}\ep_{\mu}{}^{\nu\rho}H_{\nu\rho}=-\frac{8}{5me}(\star H)_{\mu}$  
%one can obtain the answer for the full action in the new field variables which will contain also  a coupling terms involving two dual field strengts.}

To find  the full non-abelian version of the above equations  and to see how they can  be solved also with the scalar source terms present we write the field equations schematically as
\beq
\label{gaugefieldeqs}
2\ep F+m(A-B)=gXD(A,B)X,\,\,\,\,\ep G +m(A-B)=-gXD(A,B)X,
\eeq  
where all terms are gauge covariant in the broken phase ($A$ and $B$ are then the same gauge field up to covariant terms as we saw above). Solving for $B$ from the first equation and inserting the answer into the second one gives,
in the limit $g\rightarrow 0$ keeping $m=gv^2$ fixed,
\beq
\ep F=\tfrac{4}{m}\ep P(\ep F)+\tfrac{8}{m^2}\ep (\ep F,\ep F),
\eeq
where $P=\pd+ A$ and where  we have used 
\beqa
 \ep G(B)=\ep G(A+(B-A))=\ep F(A)+2\ep P(B-A)+2\ep (B-A,B-A).
 \eeqa 
 To linear order  in $\tfrac{1}{m}$  this gives the same field equation  as obtained for $C_{\mu}$ above. This may be compared to \cite{Mukhi:2011jp} where a similar set of equations is discussed. As seen there, choosing other combinations of the two gauge fields as the remaining one may lead to  situations which require 
 unlimited iterations of the kind we will see below when the scalar source terms are kept in the analysis.
 
%\footnote{This equation can be written $(1-\tfrac{4}{m}\ep P+\tfrac{8}{m^2}M)(\ep F)=0$ where $\ep P$ refers to the matrix operator $\ep_{\mu}{}^{\al\nu}(\pd+A)_{\al}$ and similarly for $M$  but with $P$ replaced by $\ep F$.} To get this result we have used the expansion
%$G_{\mu\nu}(A+a)=F_{\mu\nu}(A)+2P_{[\mu}a_{\nu]}+[a_{\mu},a_{\nu}]$ or $\ep G=\ep F+2\ep Pa+2\ep a^2$ where  $a=\ep F$.

Turning on $g$ implies that one needs to solve the equations iteratively to eliminate $B$ in the derivative  $D=\pd+A+B$ which only appears in the expression $XDX$.
This will produce an infinite series of terms in powers  of $\tfrac{1}{m}X^2$. In fact, the iteration needed is just to consider the first equation in (\ref{gaugefieldeqs}) and repeatedly eliminate  $B$ on the RHS of
 \beq
 B=A+\tfrac{2}{m}\ep F-\tfrac{g}{m}XPX-\tfrac{g}{m}X(B-A)X.
 \eeq
 Formally the solution is (for $m\neq 0$)
 \beq
 m(B-A)=\Sigma_{n=0}^{\infty}(\tfrac{X}{v})^n(2\ep F-gXPX)(\tfrac{X}{v})^n,
 \eeq
 which gives the final answer when inserted into the second field equation in (\ref{gaugefieldeqs}).

To summarize the situation in the gauge field sector: the vector fields corresponding to broken generators have all become massive in the higgs process and 
possess now one propagating mode each. 
%However, in three dimensions there are several possible  mechanisms  by means of which a 
%gauge field can pick up a mass, see, e.g., \cite{Deser:1982vy,Townsend:1983xs}.
 %The massive vector fields  found in the topologically gauged CFT realize two of these possibilities. 
 The $SO(p)$ gauge field in the final version of the theory is massive due to
 the appearance of both a Yang-Mills term and a CS term which is a generalized version  
of the higgs effect  found by Mukhi and Papageorgakis \cite{Mukhi:2008ux} (see also \cite{Mukhi:2011jp}). 
%and  used in Minkowski to provide a connection between  M2 and  D2 branes. This higgs effect  was also seen to be relevant for the
%topologically gauged ABJ(M) theories studied in \cite{Chu:2010fk}. 
%The other mechanism at work here is simpler. 
The fields $A_{\mu }^{\hat a B}$ and $B_{\mu}^{ \hat i J}$, on the other hand,  both get a mass  from a term involving the square of the gauge field which as we saw above 
gets added to their respective CS term, and  there are no Yang-Mills terms involved in these cases. In the next subsection we will identify the scalar fields
that get absorbed by the vector fields in the higgsing process.

%%%%%%%%%%%%%%%%%%%%%%%%%%%%%%%%%%%%%%%%%%%%%%%
   
\subsection{Scalar mass terms}

%%%%%%%%%%%%%%%%%%%%%%%%%%%%%%%%%%%%%%%%%%%%%%%

When we now turn to the scalar fields we need to divide them as follows: 
\beq
x^i{}_a=({\hat x}^{\hat i}{}_{\hat a}, x^{\hat i}{}_A, x^I{}_{\hat a}, x^I{}_A),
\eeq
where, since the indices $A$ and $I$ are identified,  the last field must be further split into
\beq
x^I{}_A=(z, w^{\tilde{(IA)}}, y^{[IA]}).
\eeq
Here $w$ is symmetric and traceless and $z=x^{II}=\delta^I_A\,x^I_A$. The propagating modes absorbed  by the 
gauge fields in order to become massive are $x^I{}_{\hat a}$, $x^{\hat i}{}_A$  and $y^{IJ}$, respectively, for the three mass terms in $L(v^2)$ (\ref{Lv-square}) 
discussed in the previous subsection. These three scalar fields are thus 
eliminated by the higgsing leaving only  the scalars $x^{\hat i}{}_{\hat a}, z, w^{IJ}$  in the theory. 
%In the next subsection we will obtain the mass terms  for 
%all the scalar fields and see what the symmetry breaking pattern implies in the different background geometries.

% To obtain the mass terms for the scalar fields we must expand  the potential around the VEV.  Using
%\beq
%X^2:=X^i{}_aX^{i}{}_a=(<X^i{}_a>+x^i{}_a)^2=(v\delta^I_A+x^i{}_a)^2=pv^2+2vz+x^2,
%\eeq 
%and $x^I{}_A=w^{IA}+y^{IA}+\tfrac{1}{p}\de^{IA}z$
%we  have
%\beq
%X^2=pv^2+2v\,z+x^2=pv^2+2v\,z+(\hat x)^2 + (x^{\hat i}_A)^2+(x^I_{\hat a})^2+\tfrac{z^2}{p}+y^2+w^2.
%\eeq
%
%When expanding the lagrangian below we will mostly use the $x^2$ form of the equations. 
We  need to expand the expression in (\ref{newpotential}) whose square gives the  new potential around the VEV. Using 
$X^i{}_a=v\de^I_A+x^i{}_a$
%\beq
%X^i_aX^j_a=(\tfrac{1}{\sqrt{p}}v\delta^I_A+x^i_a)(\tfrac{1}{\sqrt{p}}v\delta^J_A+x^j_a)=\tfrac{1}{p}v^2\delta^{IJ}+\tfrac{v}{\sqrt{p}}(x^i_A\delta_A^J+x^j_A\delta_A^I)
%+x^i_ax^j_a
%\eeq
%which we simply write as (making the order of the indices clear)
%\beq
%X^i{}_aX^j{}_a=v^2\delta^{IJ}+v(x^{iJ}+x^{jI})+x^i{}_ax^j{}_a,
%\eeq
%where we use an index notation that keeps track of the ranges the indices take in different terms,
we get
\begin{eqnarray}
\label{expansion}
&&X^2X^i_a-4(X_b^iX_b^j)X_a^j=(pv^2+2vz+x^2)(v\delta^I_A+x^i{}_a)
\notag\\[1mm]
&&-4(v^2\delta^{IJ}+v(x^{iJ}+x^{jI})+x^i{}_bx^j{}_b)(v\delta^J_A+x^j{}_a))\notag\\[1mm]
&&=(p-4)v^3\delta_A^I+v^2(px^i{}_a+2z\delta_A^I-4x^{Ia}-4(x^{iA}+x^{AI}))\notag\\[1mm]
&&+v(x^2\delta_A^I+2zx^i{}_a-4x^i{}_b x^A{}_b-4x^{iJ}x^{Ja}-4x^{jI}x^{ja})+x^2x^{ia}-4x^i{}_b x^j{}_b x^j{}_a.
\end{eqnarray}
The terms in the potential directly relevant for an analysis of the spectrum are of ${\mathcal O}(v^4)$. The expression that multiplies $v^4$ in the square of (\ref{expansion})
reads
%\\sextic in VEV:
%\beq
%v^6:\,\,p(p-4)^2>0,
%\eeq
%quintic in VEV\footnote{This term linear in $z$ is in the lagrangian canceled  by the  $z$-term in $X^2\bar R$ where $\bar R$ is the background value
%of the curvature scalar.}
%\beqa
%v^5:\,\,6z(p-4)^2,
%\eeqa
%quartic in VEV
\beqa
&&(3p^2-8p)x^2+(12p-64)z^2-16(p-3)x^{Ia}x^{Ia}\notag\\[1mm]
&&-16(p-3)x^{iI}x^{iI}+48x^{IJ}x^{IJ}-16(p-6)x^{IA}x^{AI}.
\eeqa

We start by analyzing the scalar fields $x^{\hat i}{}_{\hat a}$. We find
\beq
L((x^{\hat i}{}_{\hat a})^2)=-\tfrac{1}{2}(D_{\mu}x^{\hat i}{}_{\hat a})^2-\tfrac{1}{16}(x^{\hat i}{}_{\hat a})^2\bar R-\tfrac{v^4g^2}{2\cdot 32 \cdot 32}p(3p-8)(x^{\hat i}{}_{\hat a})^2.
\eeq
Inserting the constant background value for $R$, that is
\beq
\bar R=-\tfrac{6}{16 \cdot 16}g^2v^4(p-4)^2=6\Lambda(p)=-\tfrac{6}{l^2},
\eeq
gives
\beq
L((x^{\hat i}{}_{\hat a})^2)=-\tfrac{1}{2}(D_{\mu}x^{\hat i}{}_{\hat a})^2-\tfrac{2g^2v^4}{16 \cdot 16}(p-3)(x^{\hat i}{}_{\hat a})^2.
\eeq
%So the mass of $x^{\hat i}{}_{\hat a}$ can be read off as%(in both AdS and warped AdS backgrounds!):
%\beq
%{\hat m}^2(p)=\tfrac{p-3}{64}g^2v^4,
%\eeq
%which 
%in the round $AdS_3$ case obtained for $p=2$ is right on the BF bound.
% (here given with the conformal mass term included in the bound):
%\beq
%\bar R=\tfrac{2d}{d-2}\Lambda,\,\,\Lambda=-\tfrac{(d-1)(d-2)}{2l^2},\,\,m^2\geq -\tfrac{1}{l^2}\tfrac{(d-1)^2}{4}:=m^2_{(BF)},
%\eeq
%which for $d=3$ becomes
%\beq
%R=6\Lambda,\,\,\Lambda=-\tfrac{1}{l^2},\,\,m^2\geq -\tfrac{1}{l^2}=\Lambda=m^2_{(BF)}.
%\eeq
 Comparing the BF bound  to the $p=2$ scalar $x^{\hat i}{}_{\hat a}$ mass value we see that they coincide:
\beq
{\hat m}^2(p=2)=-\tfrac{1}{64}g^2v^4=\Lambda(p=2).
\eeq
%Note that with the $p$ dependence given above for ${\hat m}^2(p)$ and $\Lambda(p)$ the mass squared equal to BF bound is only true for $p=2$. 
 Also the flat Minkowski case is consistent with unitarity since 
$\hat m^2(p=4)>0$. One might also note that the two null-warped cases $p=3$ and $p=6$ with the same geometry (and perhaps without a BF-bound as 
argued in  \cite{Moroz:2009kv,Bekaert:2011cu}) seem nevertheless to be different since the masses are not the same  for the two values of $p$.
%scalars $x^{\hat i}{}_{\hat a}$ have different masses in the two cases: $\hat m^2(p=3)=0$ and $\hat m^2(p=6)=\tfrac{3}{ 64}g^2v^4$. (Thus these two cases
%behave differently with respect to the allowed range for Neumann boundary conditions.) Let us recall also that in the mass range $0\geq (ml)^2\geq-1$ scalar fields in $AdS_3$ can have either Dirichlet or Neumann boundary conditions.

We now turn to  the trace $z=x^{II}$. 
%Since  $z$ has no indices  it might  mix with some of the  gravity modes. 
%(and gauge modes if there is a $U(1)$ as e.g. for $p=2$ and $p=6$ BUT $z$  is real!??).
Using
$
X^I{}_A%=v\delta^I_A+x^I{}_A
=(v+\tfrac{z}{p})\delta^I_A+....,
$
% in the expansion of the potential term in the scalar field equation (see beginning of the previous section)
%\beq
%\Box X^i_a-\tfrac{1}{8}X^i_aR-\tfrac{g^2}{32 \cdot 32}(3X^i_a(X^2)^2-8X^i_a(X_b^jX_b^k)(X_c^jX_c^k)-16X^2X_a^kX_b^kX_b^i+48X^j_a(X_b^jX_b^k)(X_c^kX_c^i))=0,
%\eeq
%we get to linear order 
%\beqa
%&&-\tfrac{g^2}{32 \cdot 32}\delta^I_A(3\tfrac{z}{p}(v)^4p^2+3v4(v)^3\tfrac{z}{p}p^2-8\tfrac{z}{p}(v)^4p
%-8v4(v)^3\tfrac{z}{p}p)\notag\\[1mm]
%&&-16\cdot2v\tfrac{z}{p}(v)^3p-16(\tfrac{v}{\sqrt{p}})^2p3(v)^2\tfrac{z}{p}+48\cdot 5 (v)^4\tfrac{z}{p}\notag\\[1mm]
%&&=-\tfrac{15v^4g^2}{32 \cdot 32}\tfrac{z}{p}(p-4)^2,
%\eeqa
%which is the same result as obtained at order $v^4$ in the expansion of the potential above.
%In fact, the whole $z$ field equation now reads
%its field equation becomes
we get
\beq
\bar\Box z-\tfrac{1}{8}z\bar R-\tfrac{pv}{8}R^{(1)}-\tfrac{g^2v^4}{32 \cdot 32}15(p-4)^2z=0,
\eeq
where we have included the first variation of the scalar curvature in case there is a mixing between $z$ and a gravity mode. 
Inserting also the expression for the background curvature scalar $\bar R$ quoted above
%, here denoted with a bar, 
%\beq
%\bar R=-\tfrac{6}{16 \cdot 16}g^2v^4(p-4)^2,
%\eeq
%we get
%\beq
%\bar\Box z+\tfrac{1}{8}z\tfrac{6}{16 \cdot 16}g^2v^4(1-\tfrac{8}{p}+\tfrac{16}{p^2})-\tfrac{\sqrt{p}v}{8}R^{(1)}-\tfrac{v^4g^2}{32 \cdot 32}15(1-\tfrac{8}{p}+\tfrac{16}{p^2})z=0.
%\eeq
%So
it reads
\beq
\bar\Box z-\tfrac{3}{16 \cdot 16}g^2v^4(p-4)^2z-\tfrac{pv}{8}R^{(1)}=0.
\eeq
%which means that the mass square is always positive for $z$ at least before the mixing with gravity is sorted out!
To see  if there is a  mixing with gravity recall that
\beq
R^{(1)}:=\delta R=-\bar\Box h+\bar\nabla_{\mu}\bar\nabla_{\nu}h^{\mu\nu}-h^{\mu\nu}\bar R_{\mu\nu} ,
\eeq
where $\bar R_{\mu\nu}$ is the background Ricci tensor which is non-trivial in all geometries except the round  $AdS$.
 In our case we must allow 
for null-warped, and even more exotic, metrics with non-zero Cotton and traceless Ricci tensors. As we saw above, however, the scalar curvature is constant in all cases.
We will continue the analysis of the field equation for $z$ in the next subsection since we will need also  the linearized Cotton equation which is the main subject of that subsection.

Next we consider the field $w^{IJ}$ which is symmetric and traceless. We have 
%Using the expansion to the potential above we find
%\beq
%\bar\Box w-\tfrac{1}{8}w\bar R-\tfrac{g^2v^4}{32 \cdot 32}(3p^2-56p+240)w=0.
%\eeq
%or 
\beq
\bar\Box w+\tfrac{g^2v^4}{32}(p-6)w=0,
\eeq
corresponding to the mass 
\beq
\hat m^2(w)=-\tfrac{g^2v^4}{16}(p-6).
\eeq
Note that once again the null-warped cases $p=3$ and $p=6$ are different with even a zero mass value in the latter case (which also happens for  $p=3$ 
in the case of $x^{\hat i}{}_{\hat a}$). This is a property that will be significant for some of the other scalar fields in the discussion 
of the higgs effect below.

%Turning now to the scalar fields $x^{\hat i}_A$ and $x^I_{\hat a}$ which behave in a similar way with respect to the potential. From the expression 
%\beq
%3X^i_a(X^2)^2-8X^i_a(X_b^jX_b^k)(X_c^jX_c^k)-16X^2X_a^kX_b^kX_b^i+48X^j_a(X_b^jX_b^k)(X_c^kX_c^i)
%\eeq
%appearing in the Klein-Gordon equation
%%behaves very differently at the linearized level depending on which field one is considering. We have already 
%it can be seen that setting both indices $i,a$ equal to their hatted 
%values means that only the first two terms contribute. On the other hand, for the trace field $z$ all terms come in producing a complete factor $(p-4)^2$.
For  $x^{\hat i}_A$ and $x^I_{\hat a}$  we find  the same linearized field equation:
\beq
\Box x-\tfrac{1}{8}\bar R\, x-\tfrac{3}{32 \cdot 32}g^2v^4(p-4)^2x=0.
\eeq
%but now without any mixing with the gravity modes.
%Note that lineraized  Ricci tensor does not appear here since that requires looking at a field with a background value, i.e. $x^{\hat i}{}_{\hat a}$. 
Inserting the background value for the 
curvature scalar  we find for each of these fields that the total mass term vanishes for all values of $p$:
\beq
\Box x^{\hat i}_A=0,\,\,\,\Box x^I_{\hat a}=0.
\eeq
We may note that in three dimensions and  for the round $AdS_3$, this happens to be  the upper bound  of the mass, using the standard formula also for $d=3$, where both Dirichlet and Neumann boundary conditions are allowed.
%However, since the physical mass is measured relative the conformal mass term this does  not mean that the fields are massless. Instead we should write these equations as
%\beq
%\Box x-\tfrac{1}{8}\bar R\, x-M^2x=0
%\eeq
%and read off the physical mass $M$\footnote{This definition means that when massless it propagates on the light cone. It is also the mass relevant for supersymmetry, see for instance \cite{Duff:1986hr}. In particular all fields in multiplets containing an ordinary  graviton
%and/or gauge field
%are then massless.}. In these two last cases the mass square is thus
%\beq
%x^{\hat i}_A,\,\,\,x^I_{\hat a}:\,\,\,\,M^2=\tfrac{3}{32 \cdot 32}g^2v^4(p-4)^2,
%\eeq
%which is positive for all $p$ except the Minkowski case $p=4$ where it vanishes.

The final  scalar field to analyze is the anti-symmetric part of $x^{I}{}_A$. Recall the definition
%in the broken phase the indices $A$ and $I$ are identified so we should split the fluctuations $x^i_a$ into
%\beq
%x^i_a=({\hat x}^{\hat i}_{\hat a}, x^{\hat i}_A, x^I_{\hat a}, x^I_A)
%\eeq
%where the last one must be further split into (note the order of the indices: top index first, bottom one last)
$
x^I{}_A=(z, w^{\tilde{(IA)}}, y^{[IA]}),
$
with $w$ traceless and $z$  the trace of $x^I{}_A$. 
One easily checks that  the field $y^{IJ}=x^{[IJ]}$ behaves the same way as the last two scalar fields just discussed, namely
%\beq
%\Box y-\tfrac{1}{8}\bar R\, y-\tfrac{3}{32 \cdot 32}g^2v^4(1-\tfrac{4}{p})^2y=0
%\eeq
%i.e. once again we find the linear field equation after inserting the scalar curvature in any background
\beq
\Box y^{IJ}=0.
\eeq
Thus all scalar fields that are eaten by the vector fields corresponding to  broken symmetries behave this way and this is so  in all the
backgrounds discussed here. More interesting is, however, the fact  that for some  values of $p$ also  physical scalar fields behave this way. The zero mass  Klein-Gordon equation is also the equation for the singleton in $AdS_3$ \cite{Flato:1990eu}, the implications of which need further
study. However, it may be noted that in \cite{Flato:1990eu} the authors  mention two different methods to realize singletons in the $AdS_3$ bulk theory, either as vector fields or
by involving $\Box^2$ field equations. 
If and how any of these options is realized  in the present context of the topologically gauged theories considered in this paper is not clear (see, however, the next subsection). 

%%%%%%%%%%%%%%%%%%%%%%%%%%%%%%%%%%%%%%%%%%%%%%%%%%%%%%%

\subsection{Linearized field equations for maximally symmetric backgrounds $(p=2,4)$}

%%%%%%%%%%%%%%%%%%%%%%%%%%%%%%%%%%%%%%%%%%%%%%%%%%%%%%%%%

 Due to the complications in the warped cases we will in this subsection restrict ourselves to the conformally flat  cases, i.e.,  we assume that the background is either the maximally symmetric  $AdS$ or Minkowski obtained for $p=2$ and $p=4$, respectively.  We will continue to use $p$ dependent formulae when possible but we should be careful to remember  that in this subsection the results
are only valid for these two values of $p$.  

%{\it The gravitational field equation (Cotton) without matter has been discussed for the warped cases (where the critical value $\mu l=3$ gives just ordinary AdS and not the null warped case
%in Anninos et al heath 0905.2612. Note the comment on page 7 in that paper:They were not able to write the linear Cotton equation as the product of three commuting
%chiral operators as in LSS paper for ordinary AdS. The  null warped case is a pp-like limit of the warped case and needs a special metric: see Anninos et al 1005.4072.
%
% It might useful at this point to recall the metric decomposition
%\beq
%h_{\mu\nu}=h_{\mu\nu}^{TT}+ D_{(\mu}V^T_{\nu)}+(D_{\mu} D_{\nu}-\tfrac{1}{3}g_{\mu\nu}\Box)\phi+\tfrac{1}{3} g_{\mu\nu}h
%\eeq
%where $V^T_{\nu}$ is  transverse while $h_{\mu\nu}^{TT}$ is both transverse and traceless. Thus $h=\bar g^{\mu\nu}h_{\mu\nu}$.
%The six functions in $h_{\mu\nu}$ are thus  divided between the various fields in the decomposition as follows: two in the scalars $h,\phi$, 
%two  in $V^T_{\nu}$ and two in $h_{\mu\nu}^{TT}$. The two scalars can and will probably mix with the scalar $z$ since they all lack indices.
%Similar statements can be made for the vector fields in particular in cases (some $p$) where there is one or more abelian vector fields 
%massive or massless. The best chance for this to happen is in the $p=2$ case (standard AdS in fact) since then there probably is an abelian 
%$SO(p=2)$ diagonal massless vector field (which however then makes it possible to gauge away the longitudinal scalar component).}

For  maximally symmetric backgrounds  we have a zero Cotton tensor and
\beq
\bar R_{\mu\nu}=2\Lambda g_{\mu\nu}= -\tfrac{2}{16 \cdot 16}g^2v^4(p-4)^2g_{\mu\nu}.
\eeq
The first variation of the curvature scalar then becomes
\beq
R^{(1)}:=\delta R=-\bar\Box h+\bar\nabla_{\mu}\bar\nabla_{\nu}h^{\mu\nu}-h^{\mu\nu}\bar R_{\mu\nu}
=-\bar\Box h+\bar\nabla_{\mu}\bar\nabla_{\nu}h^{\mu\nu}-2\Lambda h.
\eeq
Using this expression in the  Klein-Gordon equation for the field $z$ we find
%\beq
%\bar\Box z-\tfrac{3}{16 \cdot 16}g^2v^4(p-4)^2z-\tfrac{pv}{8}(-\bar\Box h+\bar\nabla_{\mu}\bar\nabla_{\nu}h^{\mu\nu}-2\Lambda h)=0,
%\eeq
%or
%\beq
%\bar\Box z+3\Lambda z-\tfrac{pv}{8}(-\bar\Box h+\bar\nabla_{\mu}\bar\nabla_{\nu}h^{\mu\nu}-2\Lambda h)=0.
%\eeq
%In a gauge where $\bar\nabla_{\nu}h^{\mu\nu}=0$ this reads
%\beq
%\bar\Box (z+\tfrac{pv}{8}h)+\Lambda (3z+2\tfrac{pv}{8}h)=0
%\eeq
%or
%\beq
%\bar\Box (\tfrac{z}{p}+\tfrac{v}{8}h)+\Lambda (3\tfrac{z}{p}+\tfrac{v}{4}h)=0
%\eeq
%If we instead of choosing a gauge keep $\bar\nabla_{\mu}\bar\nabla_{\nu}h^{\mu\nu}=H$ we get
\beq
\bar\Box (\tfrac{z}{p}+\tfrac{v}{8}h)+\Lambda (3\tfrac{z}{p}+\tfrac{v}{4}h)=\tfrac{v}{8}H,
\eeq
where $\bar\nabla_{\mu}\bar\nabla_{\nu}h^{\mu\nu}=H$ and $h=h^{\mu}{}_{\mu}$.

We thus seem to need another equation relating the fields $z$, $h$ and $H$. This equation must come from the untraced Cotton equation
since the traced one just gives back the scalar field equation for $z$. In fact, by decomposing the metric according to 
\beq
h_{\mu\nu}=h_{\mu\nu}^{TT}+\bar D_{(\mu}V^T_{\nu)}+(\bar D_{\mu}\bar D_{\nu}-\tfrac{1}{3}\bar g_{\mu\nu}\Box)\phi+\tfrac{1}{3}\bar g_{\mu\nu}h,
\eeq
we will obtain such an equation below.
The Cotton equation is, after using the Klein-Gordon equation to eliminate some terms,
  \beqa
 &&\tfrac{1}{g}C_{\mu\nu}-\tfrac{eX^2}{16}(R_{\mu\nu}-\tfrac{1}{4}g_{\mu\nu}R)-
 \tfrac{e}{4}g_{\mu\nu}V(X)\cr
 &&-\tfrac{3e}{8}D_{\mu}X^i_aD_{\nu}X^i_a+\tfrac{e}{8}g_{\mu\nu}D^{\si}X_a^iD_{\si}X_a^i+\tfrac{e}{8}X_a^iD_{\mu}D_{\nu}X_a^i=0,
 \eeqa
 which  now has to be linearized. This has been done in many places in the literature (usually with at most  one scalar field present) and we just quote the result
  \beqa
 &&-\tfrac{v^2}{16}(\bar e\,\de^{\be}_{(\mu}-\tfrac{1}{\mu}\bar\ep_{(\mu}{}^{\al\be}\bar D_{|\al})(-\tfrac{1}{2}\bar\Box h_{\be|\nu)}+\tfrac{1}{2}\bar\nabla_{\be|}\bar\nabla^{\rho}h_{\nu)\rho}
 +\tfrac{1}{2}\bar\nabla_{\nu)}\bar\nabla^{\rho}h_{\be\rho}-\tfrac{1}{2}\bar\nabla_{\be|}\bar\nabla_{\nu)} h+\Lambda h_{\be|\nu)}-\Lambda h\bar g_{\be|\nu)})\notag\\[1mm]
&&
 +\tfrac{\bar e v\Lambda}{8}(\tfrac{z}{p}-\tfrac{v}{4}h)\bar g_{\mu\nu} +\tfrac{\bar e v^2}{64} (-\bar\Box h+H)\bar g_{\mu\nu}+\tfrac{v}{8}\bar e \bar D_{\mu}\partial_{\nu}\tfrac{z}{p}=0.
 \eeqa
 
% When analyzing this gravity equation it might be convenient to rewrite it in terms of the Lichnerowicz operator, here given for 3d maximally symmetric spacetimes as
% \beq
% \Delta_Lh_{\mu\nu}=-\bar\Box h_{\mu\nu}-2\bar R_{\mu\rho\nu\si}h^{\rho\si}+2\bar R_{(\mu}{}^{\rho}h_{\nu)\rho}=-\bar \Box h_{\mu\nu}+6\Lambda h_{\mu\nu}-2\Lambda \bar g_{\mu\nu} h,
% \eeq
%which commutes with the covariant derivative. For instance
%\beq
%D^{\nu}\Delta_Lh_{\mu\nu}= \Delta_LD^{\nu}h_{\mu\nu}.
%\eeq

If the Cotton  equation is traced we get
%\beq
%(\bar\Box +3\Lambda)\tfrac{z}{p}+\tfrac{v}{8}(\bar\Box+2\Lambda)h=\tfrac{v}{8}H,
%\eeq
%This equation should be compared to the field equation for the trace $z=x^{AA}$:
%\beq
%\bar\Box (\tfrac{z}{p}+\tfrac{v}{8}h)+\Lambda (3\tfrac{z}{p}+\tfrac{v}{4}h)=\tfrac{v}{8}H
%\eeq
%or if rearranged it a bit:
\beq
(\bar\Box +3\Lambda)\tfrac{z}{p}+\tfrac{v}{8}(\bar\Box+2\Lambda)h=\tfrac{v}{8}H
\eeq
which as expected is identical to the equation coming from the Klein-Gordon equation for $z$ given above.

We now need to analyze also the vector part of the Cotton equation. That is, we should keep the vector fields and get the equation for
$W_{\mu}=\nabla^{\mu}h_{\mu\nu}$. 
%(as in KK phys rep). 
Using the decomposition of  the metric given above
%\beq
%h_{\mu\nu}=h_{\mu\nu}^{TT}+\bar D_{(\mu}V^T_{\nu)}+(\bar D_{\mu}\bar D_{\nu}-\tfrac{1}{3}\bar g_{\mu\nu}\Box)\phi+\tfrac{1}{3}\bar g_{\mu\nu}h
%\eeq
we find after   some algebra
\beq
\tfrac{3}{2}\tfrac{\Lambda}{\mu}\ep_{\mu}{}^{\al\be}\bar D_{\al}W_{\be}=\tfrac{8}{v}\bar D_{\mu}((\bar\Box +3\Lambda)\tfrac{z}{p}+\tfrac{v}{8}(\bar\Box+2\Lambda)h-\tfrac{v}{8}H).
\eeq
The scalar equation for $z$ obtained above puts the expression in the RHS bracket to zero and hence
\beq
\ep_{\mu}{}^{\al\be}\bar D_{\al}W_{\be}=0.
\eeq
%where we used the notation 
%\beq
%W_{\mu}=D^{\nu}h_{\mu\nu}.
%\eeq
Relating $V_{\mu}^T$ in the metric decomposition to $W_{\mu}$ we get an equation whose divergence becomes
%\beq
%W_{\mu}=\tfrac{1}{2}(\Box+2\Lambda)V_{\mu}^T+\tfrac{2}{3}D_{\mu}(\Box+3\Lambda)\phi+\tfrac{1}{3}D_{\mu}h,
%\eeq
%which can be written in terms of the spin 1 Hodge-de Rham operator as
%\beq
%W_{\mu}=\tfrac{1}{2}(-\Delta_1+4\Lambda)V_{\mu}^T+\tfrac{2}{3}D_{\mu}(\Box+3\Lambda)\phi+\tfrac{1}{3}D_{\mu}h,
%\eeq
%and hence its divergence is
\beq
H=\tfrac{2}{3}\bar\Box(\bar\Box+3\Lambda)\phi+\tfrac{1}{3}\bar\Box h,
\eeq
and using this  in the scalar equation for $z$ leads to the following result:
%\beq
%\bar\Box (\tfrac{z}{p}+\tfrac{v}{8}h)+\Lambda (3\tfrac{z}{p}+\tfrac{v}{4}h)=\tfrac{v}{8}( \tfrac{2}{3}\Box(\Box+3\Lambda)\phi+\tfrac{1}{3}\Box h),
%\eeq
%giving
%\beq
%(\bar\Box +3\Lambda)(\tfrac{z}{p}+\tfrac{v}{12}h)=\tfrac{v}{12}\Box(\Box+3\Lambda)\phi,
%\eeq
%or simply
\beq
(\bar\Box +3\Lambda)(\tfrac{z}{p}+\tfrac{v}{12}(h-\bar\Box\phi))=0,
\eeq
which means that there is actually only one physical index-free 
scalar field in the theory.

In order to choose a convenient gauge\footnote{Fixing the gauge completely, e.g.,  using   the physical light-cone gauge as done in \cite{Nilsson:2008ri}, one finds that all non-zero components of the metric can be expressed  in terms of the stress tensor for the matter fields.} we note that the  equation $\ep_{\mu}{}^{\al\be}\bar D_{\al}W_{\be}=0$  
%\beq
%\ep_{\mu}{}^{\al\be}D_{\al}(\Box+2\Lambda)V^T_{\be}=0,
%\eeq
%which means that either $V^T_{\mu}=0$ or a mode satisfying $(\Box+2\Lambda)V^T_{\be}=0$.
suggests the gauge choice $V^T_{\mu}=0$. Choosing also $\bar \Box\phi=h$ 
%($h_{\mu\nu}$ still massless so OK ???????) 
it follows  that
\beq
h_{\mu\nu}=h_{\mu\nu}^{TT}+(\bar D_{\mu}\bar D_{\nu}-\tfrac{1}{3}\bar g_{\mu\nu}\bar \Box)\phi+\tfrac{1}{3}\bar g_{\mu\nu}h=
h_{\mu\nu}^{TT}+\bar D_{\mu}\bar D_{\nu}\phi,
\eeq
and
 \beq
H=\bar D^{\mu}\bar D^{\nu}\bar D_{\mu}\bar D_{\nu}\phi=(\bar\Box+2\Lambda)\bar\Box\phi=(\bar\Box+2\Lambda) h.
\eeq
As in the previous subsection,  we find also here  some features  indicating that  $AdS_3$ bulk singletons play a role. Writing the parameter of coordinate transformations as $\xi_{\mu}=\xi^T_{\mu}+\pd_{\mu}\xi$
we get a transformation of the trace of the metric involving a $\Box$ which together with the appearance of $\Box^2$ above should be compared to the discussion in  \cite{Flato:1990eu}.

Finally, the equation for the traceless transverse part of the metric $h_{\mu\nu}^{TT}$ is identical to the one obtained in pure TMG \cite{Li:2008dq} namely
\beq
({\cal \bar D}(\mu){\cal \bar D}(l){\cal \bar D}(-l))_{(\mu}{}^{\rho}h_{\nu)\rho}^{TT}=0,
\eeq
where the operators $\mathcal D(l)$ etc are defined as
\beq
{\cal \bar D}(l)_{\mu}{}^{\rho}=\bar e\de^{\rho}_{\mu}-\tfrac{1}{l}\bar \ep_{\mu}{}^{\al\rho}\bar D_{\al}.
\eeq

An analysis with more properties of  supergravity   taken into account can be found in the 
work by Becker et al. \cite{Becker:2009mk}. In particular, it is found there that at the critical point (and only there) super-TMG theories
with ${\cal N}=(1,0)$, $ {\cal N}=(0,1)$ and  ${\cal N}=(1,1)$ supersymmetry but without a matter sector  satisfy a positive  energy theorem (in the sign conventions of \cite{Li:2008dq})
and are chiral in the same sense as in the bosonic case studied in \cite{Li:2008dq}.

%%%%%%%%%%%%%%%%%%%%%%%%%%%%%%%%%%%%%%%%%%%%%%

%\subsection{Supersymmetry: local and rigid}

%%%%%%%%%%%%%%%%%%%%%%%%%%%%%%%%%%%%%%%%%%%%%%

%%%%%%%%%%%%%%%%%%%%%%%%%%%%%%%%%%%%%%%%%%%%%%

\subsection{Susy rules for any $p$}

%%%%%%%%%%%%%%%%%%%%%%%%%%%%%%%%%%%%%%%%%%%%%%
In this subsection we will briefly discuss what the transformation rules tell us about the multiplet structure in the different backgrounds.
The following formulae are valid for all values of $p$.
%It will then also be appropriate to check how the theory behaves when certain limits in the parameters are taken. 
%In particular we will check if we can obtain theories with rigid supersymmetry in curved backgrounds by decoupling gravity. 
%As we will see, however, contrary to the ABJ(M) case studied in \cite{Chu:2010fk}, this option will not be present in the $SO(N)$ theory
%analyzed here. For this to be possible the BLG coupling $\lambda$ must be  reinstating as will be briefly discussed in the conclusions.
%

%Before starting the analysis of the transformation rules it would be good to have some feeling for the structure of the supersymmetry parameter after symmetry breaking.
%In the Minkowski case (p=4) we have ($\ep$ is a spinor under $SO(8)$ R-symmetry and $\psi_a$ is a co-spinor, up to triality)
%\beq
%SO(8)\rightarrow SO_1(4)\times SO_2(4):\,\,v\rightarrow(4,1)+(1,4),\,\,s\rightarrow (s,s)+(c,c),\,\,\,c\rightarrow (s,c)+(c,s)
%\eeq
%as seen from using the basis (1,0,0,0) etc for v and (1/2,1/2,1/2,1/2) etc for s and c. Similarly for maximally symmetric AdS solution (p=2) we get
%\beq
%SO(8)\rightarrow SO(6)\times SO(2):\,\,v\rightarrow 6_0+1_1+1_{-1},\,\,s\rightarrow 4_1+\bar 4_{-1},\,\,\,c\rightarrow 4_{-1}+\bar 4_{1}
%\eeq

The fields that appear after the superconformal symmetry breaking will organize themselves into supermultiplets according to their number of $SO(N-p)$
vector indices for the simple  reason  that the supersymmetry parameter does not have any such indices. Thus we find one multiplet with $8(N-p)$ d.o.f. for both bosons and fermions
containing the following fields (the $\hat a$-vector multiplet)
\beq
x^{\hat i}{}_{\hat a},\,\,\psi_{\hat a},\,\,\,A_{\mu}^{A\hat a}(massive),
\eeq
and one with $8p$ d.o.f. for both bosons and fermions
containing  (the $\hat a$-scalar multiplet)
\beq
C_{\mu}{}^{IJ}{}(massive),\,\,w^{IJ},\,\,z,\,\,\,\psi_{ A},\,\,\,B_{\mu}^{\hat i J}(massive).
\eeq
%which together account for all gauge field components as well as the $8N+8N$ physical degrees of freedom. This structure is directly suggested by the transformation rule for the scalar fields
%\beq
%\de X^i_a=i\bar\ep_m\Ga^i\psi_a.
%\eeq
The remaining vector  fields  $A_{\mu}^{\hat a \hat b}(massless)$ and $B_{\mu}^{\hat i \hat j}(massless)$ couple to both multiplets  as usual for CS gauge fields carrying no degrees of freedom.
These two multiplets will also couple to the gravitational field 
with spin 2  which is still massless and without propagating degrees of freedom. The corresponding statement for the spin 3/2 fields depends on the number of 
surviving supersymmetries\footnote{This number will depend on $p$ as is clear from the analysis of \cite{Gibbons:2008vi}.}. Below we will present some properties of the transformation rules that support this picture.

The  supersymmetry transformation rules are as quoted from  \cite{Gran:2008qx,Gran:2012mg}, with $\ep_m=A\ep_g$ and $A^2=\thalf$, 
% \beq
%\delta e_{\mu}{}^{\alpha}=i\bar\epsilon_g \gamma^{\alpha}\chi_{\mu},
%\eeq
%\beq
% \delta\chi_{\mu}= \tilde D_{\mu}\epsilon_g.
%\eeq
%\beqa
%\de B_{\mu}^{ij}&=&-\tfrac{i}{2e}\bar \ep_g\Ga^{ij}\ga_ {\nu}\ga_{\mu}f^{\nu}-\tfrac{3ig}{8}\bar\psi_a\ga_{\mu}\Ga^{[i}\ep_mX^{j]}_a
%-\tfrac{ig}{16}\bar\psi_a\ga_{\mu}\Ga^{ijk}\ep_mX^{k}_a\notag\\[1mm]
%&&-\tfrac{ig}{4}\bar\chi_{\mu}\Ga^{k[i}\ep_gX^{j]}_aX^k_a-\tfrac{ig}{32}\bar\chi_{\mu}\Ga^{ij}\ep_gX^2
%\eeqa
%\beq
%\de X^i_a=i\bar\ep_m\Ga^i\psi_a
%\eeq
%  \beqa
%  \de \psi_a&=&\ga^{\mu}\Ga^i\ep_m(\tilde D_{\mu}X^i_a-iA\bar\chi_{\mu}\Ga^i\psi_a)+\tfrac{\lambda}{6}\Ga^{ijk}\ep_mX^i_bX^j_cX^k_d\,f^{bcd}{}_a\notag\\[1mm]
%  &&+\tfrac{g}{8}\Ga^i\ep_mX^i_bX^j_bX^j_a-\tfrac{g}{32}\Ga^i\ep_mX^i_aX^2
%    \eeqa
%  \beqa
%\de \tilde A_{\mu}^{ab}&=&-i\lambda\bar\ep_m\ga_{\mu}\Ga^i\psi_cX^i_d\,f^{cdab}-\tfrac{i\lambda}{2}\bar\chi_{\mu}\Ga^{ij}\ep_gX^i_cX^j_d\,f^{cdab}\notag\\[1mm]
%&&+\tfrac{ig}{4}\bar\ep_m\ga_{\mu}\Ga^i\psi_{[a}X^i_{b]}+\tfrac{ig}{8}\bar\chi_{\mu}\Ga^{ij}\ep_gX^i_aX^j_b
%\eeqa
%
%We will consider these with the BLG coupling $\lambda=0$:
 \beq
\delta e_{\mu}{}^{\alpha}=i\bar\epsilon_g \gamma^{\alpha}\chi_{\mu},
\eeq
\beq
 \delta\chi_{\mu}= \tilde D_{\mu}\epsilon_g,
\eeq
\beqa
\de B_{\mu}^{ij}&=&-\tfrac{i}{2e}\bar \ep_g\Ga^{ij}\ga_ {\nu}\ga_{\mu}f^{\nu}-\tfrac{3ig}{8}\bar\psi_a\ga_{\mu}\Ga^{[i}\ep_mX^{j]}_a
-\tfrac{ig}{16}\bar\psi_a\ga_{\mu}\Ga^{ijk}\ep_mX^{k}_a\notag\\[1mm]
&&-\tfrac{ig}{4}\bar\chi_{\mu}\Ga^{k[i}\ep_gX^{j]}_aX^k_a-\tfrac{ig}{32}\bar\chi_{\mu}\Ga^{ij}\ep_gX^2,
\eeqa
\beq
\de X^i_a=i\bar\ep_m\Ga^i\psi_a,
\eeq
  \beqa
  \de \psi_a&=&\ga^{\mu}\Ga^i\ep_m(\tilde D_{\mu}X^i_a-iA\bar\chi_{\mu}\Ga^i\psi_a)\
  +\tfrac{g}{8}\Ga^i\ep_mX^i_bX^j_bX^j_a-\tfrac{g}{32}\Ga^i\ep_mX^i_aX^2,
    \eeqa
  \beqa
\de A_{\mu}^{ab}&=&\tfrac{ig}{4}\bar\ep_m\ga_{\mu}\Ga^i\psi_{[a}X^i_{b]}+\tfrac{ig}{8}\bar\chi_{\mu}\Ga^{ij}\ep_gX^i_aX^j_b,
\eeqa
%Here we should note that the last transformation rule is  non-linear and relate fields with different gauge index structure. In the analysis
%below, we will see that in the various backgrounds this feature is  present only for the massless gauge fields while, due to the linearization around the VEV's of the scalar fields, the massive gauge fields will transform under supersymmetry with linear transformation rules.
which we want to  linearize  around a general background. Consider first $\de \psi_a$ written as
\beq
\de \psi_a=\ga^{\mu}\Ga^i\ep_m(\tilde D_{\mu}X^i_a-iA\bar\chi_{\mu}\Ga^i\psi_a)\
  -\tfrac{g}{32}\Ga^i\ep_m(X^2X^i_a-4X^i_bX^j_bX^j_a),
\eeq
where we recognize in the last term the expression whose square is the potential and which has been expanded in powers of the VEV in the previous section.
%Multiplied by $\Ga^i\ep_m$ this expression becomes
%\beq
%\Ga^i\ep_m(X^2X^i_a-4X^i_bX^j_bX^j_a)=\Ga^{\hat i}\ep_m(X^2X^{\hat i}_a-4X^{\hat i}_bX^j_bX^j_a)+\Ga^I\ep_m(X^2X^I_a-4X^I_bX^j_bX^j_a).
%\eeq
%Then considering the two types of 3-algebra indices gives: first we set $a=A$
%\beqa
%&&\Ga^I\ep_m(X^I_bX^j_bX^j_A-\tfrac{1}{4}X^I_AX^2)+\Ga^{\hat i}\ep_m(X^{\hat i}_bX^j_bX^j_A-\tfrac{1}{4}X^{\hat i}_AX^2)\notag\\[1mm]
%&&=\Ga^I\ep_m(\tfrac{v^3}{p\sqrt{p}}\de^I_A+\tfrac{v^2}{p}(x^{IA}+x^{AI}+x^{IA}))-\tfrac{1}{4}(\tfrac{v^3}{\sqrt{p}}\de^I_A+v^2x^I_A+\tfrac{2v^2}{p}\de^I_Az))\notag\\[1mm]
%&&+\Ga^{\hat i}\ep_m(\tfrac{v^2}{p}x^{\hat i}{}_A-\tfrac{v^2}{4}x^{\hat i}{}_A)+{\it O}(x^2)\notag\\[1mm]
%\eeqa
%where we used
%\beq
%X^i_a=\tfrac{v}{\sqrt{p}}\de^I_A+x^i_a,\,\,\,(X^i_a)^2=v^2+\tfrac{2v}{\sqrt{p}}z+(x^i_a)^2
%\eeq
%Summing up this general case we find

Choosing first $a=A$ we get
\beqa
\de\psi_A&=&\ga^{\mu}\Ga^{\hat i}\ep_m D_{\mu}X^{\hat i}{}_A+\ga^{\mu}\Ga^{I}\ep_m D_{\mu}X^{I}{}_A-\tfrac{gv^3}{32}\Ga^I\ep_m(p-4)\de^I_A\notag\\[1mm]&&+\tfrac{gv^2}{8}\Ga^{I}\ep_m(x^{IA}+2x^{(IA)})-\tfrac{gv^2}{32}\Ga^{I}\ep_m(p\,x^I{}_A+2\de^I_Az)\notag\\[1mm]
&&-\tfrac{gv^2}{32}\Ga^{\hat i}\ep_m(p-4)x^{\hat i A}+{\it O}(x^2).
\eeqa

Note that in all non-Minkowskian backgrounds ($p\neq 4$) there are non-zero constant terms % and blows up in the gravity decoupling limit $v\rightarrow \infty$ (with $gv^2$ fixed, or rather the geometry ie $gv^2\propto 1/l$ fixed and let $g\rightarrow 0$). 
indicating a symmetry breaking of the superconformal symmetry. However, these terms can  be removed by adding a special superconformal transformation 
\beq
\de_S\psi_a=<X^i{}_a>\Ga^i\eta_m,\,\,\,\eta_m=\ep_m\tfrac{gv^2}{32}(p-4).
\eeq
Thus, the
$Q$ transformations present in any of the broken  phases (except the Minkowski one) are obtained by this special  combination of  the $Q$ and $S$  transformations
in the  unbroken conformal phase. 
%arising for $p=4$.  This is just as we would expect since
%the covariant derivative in $\de_Q\chi_{\mu}|_{AdS}=\de_Q\chi_{\mu}|_{conf}+\de_S\chi_{\mu}|_{conf}$ does not play a role in the flat case.
For instance, in the round AdS case obtained for $p=2$ this leads to the covariant derivative
%\footnote{Note that for dimensional reasons the special superconformal transformations will not be altered
%by the topological gauging.}
\beq
\de\chi_{\mu}=D_{\mu}\ep_g+\ga_{\mu}\eta_g=(D_{\mu}-\tfrac{gv^2}{16}\ga_{\mu})\ep_g:=\hat D_{\mu}\ep_g,
\eeq
where we assumed that the same relation between $\eta_m$ and $\eta_g$ is true as for the ordinary susy parameters. Note that as expected the new term is related to the cosmological constant 
\beq
\Lambda=-\tfrac{1}{l^2}=-\tfrac{g^2v^4}{16\cdot 16}(p-4)^2,
\eeq
as $\tfrac{gv^2}{16}=\tfrac{1}{2l}$.
Thus for $p=2$ we find that
\beq
\hat D_{\mu}=D_{\mu}-\tfrac{1}{2l}\gamma_{\mu},
\eeq
which  is the same result as found in the ABJM case in \cite{Chu:2010fk}. %See also Bergshoeff et al.....
In fact, this form of the covariant derivative is valid for all values of $p$ (with $l=\infty$ for $p=4$).
%However, note that there is a sign issue here since the relation that is being used  is $\tfrac{gv^2}{16}(p-4)=\pm \tfrac{1}{2l}$.

With this understanding of the mixing of Q and S transformations we have
%\beqa
%\de\psi_A&=&\ga^{\mu}\Ga^{\hat i}\ep_m(\partial_{\mu}x^{\hat i}{}_A+vB_{\mu}^{\hat i A})+\ga^{\mu}\Ga^{I}\ep_m(\partial_{\mu}x^{I}{}_A+
%v(B_{\mu}^{IA}+A_{\mu}^{AI}))
%\notag\\[1mm]
%&&+\tfrac{gv^2}{8}\Ga^{I}\ep_m(x^{IA}+2x^{(IA)})-\tfrac{gv^2}{32}\Ga^{I}\ep_m(p\,x^I{}_A+2\de^I_Az)\notag\\[1mm]
%&&-\tfrac{gv^2}{32}\Ga^{\hat i}\ep_m(p-4)x^{\hat i A}+{\it O}(x^2),
%\eeqa
%where we also linearized the connection terms. Here the first bracket shows that the broken gauge field $B_{\mu}^{\hat i A}$ will absorb $\partial_{\mu}x^{\hat i}{}_A$ and form
%the massive new vector field  with the same name. This is just a gauge transformation using the broken symmetry parameter $\Lambda^{\hat i}{}_A$. Following the standard
%procedure the field $x^{\hat i}{}_A$ can then be gauged away and hence set to zero everywhere. The same applies to $x^{[IJ]}$ and $A_{\mu}^{IA}-B_{\mu}^{IA}$ which is the combination of these gauge fields that  picks a mass term from the Klein-Gordon term. Thus 
%in the broken phases we should write this transformation rule as follows
\beqa
\de\psi_A&=&\ga^{\mu}\Ga^{I}\ep_mD'_{\mu}(w^{IA}+\tfrac{1}{p}\de^{IA}z)-
v(A_{\mu}^{IA}-B_{\mu}^{IA}))
+\ga^{\mu}\Ga^{\hat i}\ep_m vB_{\mu}^{\hat i A}\notag\\[1mm]
&&+\tfrac{gv^2}{8}\Ga^{I}\ep_m(3w^{IA}+\tfrac{3}{p}\de^{IA}z)-\tfrac{gv^2}{32}\Ga^{I}\ep_m(p\,w^{IA}+3\de^I_Az)]+{\it O}(x^2),
\eeqa 
where $D'_{\mu}x^{I}{}_{A}=\pd_{\mu}x^{I}{}_{A}+A_{\mu AB}x^I{}_B+B_{\mu}^{IJ}x^J{}_A$ and where we have only kept the physical scalar fields that are not eaten in the higgs effect. At this point we should recall the discussion in the beginning of this section concerning the reduction of the two gauge fields to a single massive one and the 
structure of the interaction terms involving the remaining scalar fields that arose in that analysis. Using that information we will find that the above transformation rule is in fact rather non-trivial
when written out in detail.
%Note that in the Minkowski case $p=4$ this simplifies and the last line above becomes
%\beqa
%&&=\Ga^I\ep_m(\tfrac{v^2}{8}(x^{IA}+x^{AI}+x^{IA}))-\tfrac{1}{8}(v^2x^I{}_A+\tfrac{2v^2}{4}\de^I_Az))+{\it O}(x^2)\notag\\[1mm]
%%&&=\Ga^I\ep_m\tfrac{v^2}{8}(x^{IA}+x^{AI}-\tfrac{1}{2}\de^I_Az)\notag\\[1mm]
%%&&=\Ga^I\ep_m\tfrac{v^2}{8}(2w^{IA}+\tfrac{2}{4}\de^{IA}z-\tfrac{1}{2}\de^I_Az)
%&&=\Ga^I\ep_m\tfrac{v^2}{4}w^{IA}+{\it O}(x^2).
%\eeqa
%In fact, one can check that in the Minkowski background the scalar field $z$ cancels out to all orders in this expression and therefore only appears
%in $\de\psi_A$ linearly in a derivative term.
%where the decomposition has been used
%\beq
%x^{IA}=y^{[IA]}+w^{(IA)}+\tfrac{1}{p}\de^{IA}z
%\eeq
%
%Thus summing the situation for the Minkowski case:
%\beqa
%\de\psi_A&=&\ga^{\mu}(\Ga^{\hat i}\ep_m\tilde D_{\mu}x^{\hat i}{}_A+\Ga^{I}\ep_m\tilde D_{\mu}X^{IA})
%+\tfrac{v^2g}{16}\Ga^I\ep_mw^{IA}+{\it O}(x^2)\notag\\[1mm]
%&&=\ga^{\mu}(\Ga^{\hat i}\ep_m(\partial_{\mu}x^{\hat i}{}_A+\tfrac{v}{\sqrt{p}}B_{\mu}^{\hat i J}\de^J_A)+\Ga^{I}\ep_m(\partial_{\mu}X^{IA}+\tfrac{v}{\sqrt{p}}A_{\mu AB}\de_B^I+\tfrac{v}{\sqrt{p}}B_{\mu}^{IJ}\de_ A^J))\notag\\[1mm]
%&&+\tfrac{v^2g}{16}\Ga^I\ep_mw^{IA}+{\it O}(x^2)
%\eeqa

Next we consider the transformation rule for the other choice of index, i.e.,  $a={\hat a}$, 
%\beqa
%&&\Ga^I\ep_m(X^I{}_{\hat a}X^2-4X^I_bX^j_bX^j{}_{\hat a})+\Ga^{\hat i}\ep_m(X^{\hat i}{}_{\hat a}X^2-4X^{\hat i}_bX^j_bX^j{}_{\hat a})\notag\\[1mm]
%&&=\Ga^I\ep_mv^2(p-4)x^{I\hat a}+{\it O}(x^2),
%\eeqa
%which vanishes in the Minkowski case. Thus the index choice $a={\hat a}$ gives
%\beqa
%\de\psi_{\hat a}&=&\ga^{\mu}(\Ga^{\hat i}\ep_m \partial_{\mu}x^{\hat i\hat a}+\Ga^{I}\ep_m(\partial_{\mu}x^{I\hat a}+A_{\mu}^{\hat a B}v\de_B^I))\notag\\[1mm]
%&&-\tfrac{gv^2}{32}(p-4)\Ga^I\ep_mx^{I\hat a}+{\it O}(x^2),
%\eeqa
which after higgsing reads
\beqa
\de\psi_{\hat a}&=&\ga^{\mu}(\Ga^{\hat i}\ep_m D'_{\mu}x^{\hat i\hat a}+v\Ga^{I}\ep_mA_{\mu}^{\hat a I})-\tfrac{gv^2p}{32}\Ga^{\hat i}\ep_m\hat x^{\hat i}{}_{\hat a}+{\cal O}(x^2),
%-\tfrac{gv^2}{32}(p-4)\Ga^I\ep_mx^{I\hat a}
%+{\it O}(x^2)
\eeqa
which  also supports the multiplet structure given above.
There are many  features here that need further study. These will be studied    elsewhere.
%
%%%%%%%%%%%%%%%%%%%%%%%%%%%%%%%%%%%%%%%%%%%%%%%%%%%%%

%\subsubsection{The decoupling limit and rigid susy in non-trivial geometries}

%%%%%%%%%%%%%%%%%%%%%%%%%%%%%%%%%%%%%%%%%%%%%%%

%%%%%%%%%%%%%%%%%%%%%%%%%%%%%%%%%%%%%%%%%%%%%%%%%%%%%

%\section{Comments on topologically gauged ABJ(M)}

%%%%%%%%%%%%%%%%%%%%%%%%%%%%%%%%%%%%%%%%%%%%%%%

%In the topologically gauged ABJ(M) theory there are besides $N,N'$ three other parameters to play with: $\lambda, g, v$. This makes it  possible
%to define different kinds of limits  as done in \cite{Chu:2010fk}. In particular a rigid susy possibility arises as follows:\\
%?????????????????????????????\\
%?????????????????????????????\\
%

%%%%%%%%%%%%%%%%%%%%%%%%%%%%%%%%%%%%%%%%%%%%%%%%%%%%%%%%

\section{Comments}

%%%%%%%%%%%%%%%%%%%%%%%%%%%%%%%%%%%%%%%%%%%%%%%%%%%%%%%%

 The topologically gauging \cite{Gran:2008qx} of free  matter CFTs  in three dimensions with eight supersymmetries gives rise to an $O(N)$ type  model
 with a novel six order scalar potential \cite{Gran:2012mg}. This potential  consists of three different triple trace terms, one of them being  $(X^2)^3$ where $X^2:=X^i{}_aX^i{}_a$ with $i=1,2,..,8$ and $a=1,2,..,N$. Neglecting the R-symmetry index this term is precisely 
 the scalar term $(\phi^i\phi^i)^3$ that has been discussed recently, see, e.g., Aharony et al.
 \cite{Aharony:2011jz}, in the context of the $AdS_4/CFT_3$ correspondence  relating   $O(N)$ models in three dimensions to  four-dimensional Vasiliev higher spins systems \cite{Vasiliev:1995dn}. Note that when construction the other two sixth order terms appearing in the potential of this topologically gauged model 
 the R-symmetry index play a key role. These two terms in the potential are therefore not present in the  usual treatments of marginal deformations of   $O(N)$ type  models in three dimensions  but  are crucial for the critical solutions to appear in our models. 
 
 In relation to the $AdS_4/CFT_3$ correspondence it may also the pointed out that in these topologically gauged  $O(N)$ models the Chern-Simons terms of both the vector fields and the spin connection are multiplied by the same coupling constant (denoted $g$). Thus if the interpolation between the $A$ and $B$ type HS models in \cite{Vasiliev:1995dn}, parametrized by the parameter $\theta_0$,  is related to
 the introduction of gauge interactions and a non-parity symmetric Chern-Simons term as argued in \cite{Giombi:2012ms}, and hence also to the related bosonization phenomenon,  then in versions with ${\mathcal N}=8$ supersymmetry also gravitational Chern-Simons terms will enter on the field theory side. One may speculate that
 such spin two terms may be related to turning on $\theta_2$, the second coefficient among the $\theta_{2n}$ parameters defining the HS theories that interpolate 
 between the $A$ and $B$ type models in Vasiliev's system in $AdS_4$. 
 
Some features of  topologically gauged CFTs  indicate  that they may have a deeper role to play in the context of  $AdS/CFT$. The $AdS_4/CFT_3$ correspondence was mentioned above but also  the $AdS_3/CFT_2$ correspondence has recently  been investigated in depth in many papers using $W_N$ algebras in two dimensions and its connection to Vasiliev's higher spin systems in three dimensions, see \cite{Gaberdiel:2010ar}.
 In view of the fact that  $AdS_3$  arises naturally as a spontaneously broken phase of a three-dimensional topologically gauged superconformal theory as discussed in this paper, one may ask if this conformal theory
 could not itself be the boundary theory of an $AdS_4$  theory.  The  {\it sequential AdS/CFT} that is suggested  by these facts was first discussed  
  in \cite{Nilsson:2012ky}. The new information since that paper was written, namely  that the topologically gauged $CFT_3$ with eight supersymmetries is 
 actually a kind of  $O(N)$ model,  may thus be important. Also the possible role of singletons  found in this paper may be pointing in the direction 
 of such a {\it sequence}. In the topologically gauged ABJ(M) models first derived in \cite{Chu:2009gi} and developed further in \cite{Chu:2010fk} the situation is a bit more complicated since in that case there are more than one  independent coupling constant for any choice of  gauge group. The idea that several $AdS/CFT$s may follow one after the other has appeared previously in the literature. Based on higher spin and unfolding arguments, Vasiliev  raised this possibility in \cite{Vasiliev:2001zy}
 and made it explicit in a recent paper \cite{Vasiliev:2012vf}. Speculations with the same goal based on $AdS_d$ foliations of $AdS_{d+1}$ can be found in 
 \cite{Compere:2008us} (see also \cite{Aharony:2010ay} for related comments).
 However, the scenario of a "sequential AdS/CFT" coming from a topologically gauged $CFT_3$ is the first one which relies on a dynamical model
 and a conformal symmetry breaking mechanism interpolating between two $AdS/CFT$s as pointed out in \cite{Nilsson:2012ky}.

 The main purpose of this paper was to  elaborate on the observation  that the topologically gauged  $O(N)$ theory with eight supersymmetries  has a number of  special
 background solutions with interesting properties. These solutions,  of which two   were found in \cite{Gran:2012mg}, depend on the number of scalar fields that are given a VEV and can be characterized 
 by the value of $\mu l$ where $\mu$ is the coupling constant of the  gravitational CS term and $l$ is related as usual to the cosmological constant.
 The solutions that appear  correspond to the values $\mu l= \tfrac{1}{3},1,3,\infty,5,3,\tfrac{7}{3},2$. Here we recognize the second one as connected to chiral gravity,
 the third and sixth ones to the null-warped, or Schr\"odingier($z=2$), geometry while $\mu l=5$ can be associated with a solution recently discovered in \cite{Ertl:2010dh}.
 $\mu l=\infty$ corresponds to Minkowski space and requires a separate discussion.
  
 In this paper we have tried to argue that although for each of these values there are more than one kind of solution, the ones that are relevant as broken phases of the superconformal
 topologically gauged theory  are only the "critical" ones\footnote{A way to make this more concrete may be to consider the unitary representations of $SO(3,2)$ that are involved and how they behave under the symmetry breaking.  This way of looking at it could, e.g., explain why only the  
 representations of  $SO(2,2)$ with the correct properties 
 appear in $AdS_3$.}. For $\mu l=1$ this is based on the fact that the critical, or chiral,  case has no propagating massive gravitons
 which should be a direct consequence of the connection to the superconformal unbroken phase which is also lacking such modes. The $\mu l=3$ null-warped, or Schr\"odingier($z=2$), case
 has also been argued to be chiral in \cite{Anninos:2010pm} but is also "critical" for seemingly different reasons, see, e.g., \cite{Blau:2009gd}. The working hypothesis adopted here  that all the above values of $\mu l$ have  special solutions
  is indeed also supported by the existence of a special solution for  $\mu l=5$ \cite{Ertl:2010dh}. 
 The topologically gauged ABJ(M) theory \cite{Chu:2009gi} have similar properties  but for a smaller set of solutions.
 
 For $p=8$ we get $\mu l=2$ which stands out because it is even. If there is a special solution of this kind it should contain odd 
 powers\footnote{See {\bf Note added} at the end of this section for recent developments.}  of $e^{\rho/l}$. Examples with such 
 a dependence on $\rho$ are known in theories containing  a scalar field with a potential, see, e.g., \cite{Berg:2001ty}. In \cite{Cunliff:2013en} the Fefferman-Graham expansion for NMG is discussed in detail and  a generalized expansion  introduced that can accommodate both novel boundary behavior in $AdS$ as well as entirely different non-$AdS$ boundary behavior like for the $\mu l=3$ null-warped solution.  There are also  generalizations  with
 higher values $\mu l=5,7,...$  \cite{Ahmedov:2010em, Ahmedov:2010uk}. 
  
 The "critical"  null-warped, or Schr\"odinger($z=2$), solution is one of the most attractive three-dimensional geometries for condensed matter applications.
 This geometry (often with extra flat directions) is designed to have  Schr\"odinger symmetries on the boundary that play a crucial role 
 in, e.g.,  unitary Fermi gases (cold atoms) etc.

Finally, let us return to the topologically gauged ABJ(M) case mentioned in the introduction. There we recalled the result from \cite{Chu:2009gi} 
that giving a real VEV $v$ to
one of the complex scalar fields gives rise to a background solution corresponding to a  super-TMG theory at the chiral point. In the context of the 
topologically gauged $SO(N)$ model investigated in this paper the VEV was generalized to a $p\times p$ diagonal VEV matrix  leading to a number of interesting
backgrounds. Repeating this step for the ABJ(M) case we find, for $p\leq 4$,
\beq
\mu l=\sqrt{\frac{-3p^2}{5p^2-24p+16}}.
\eeq 
The values produced by this formula are
\beq
\mu l=1,1,\sqrt{\tfrac{27}{11}}, \infty,
\eeq
where we recognize the first two as critical round $AdS$ and the last one as Minkowski.
This analysis for ABJ(M) is valid for infinite level but one should note that if the other two sets of potential terms (i.e., the single and double trace terms) are kept they may  be non-zero in some of these backgrounds.
From the properties of the structure constants $f^{ab}{}_{cd}$ summarized in \cite{Nilsson:2012ky} we see that even the part of the potential linear in the structure
constant may contribute: $f^{ab}{}_{ab}\rightarrow N^2N'-NN'^2$ giving $p(p-1)$ in a vector model (that is, for $N'=1$) with $N=p$  in the background.\\
\\
\noindent{\bf Note added.}

Since this paper appeared on the ArXiv, there has been developments relevant  for the list of known solutions realizing  the values of $\mu l$ listed in  (\ref{mu-l-values}) and discussed in section 2.3. There the value $\mu l=2$ was not discussed since no such solution seemed to be known in the literature. However, a solution with $\mu l=2$ was found recently in extended topologically massive gravity with $(1,1)$ supersymmetry in \cite{Deger:2013yla}.  
That solution involves  in a crucial way a topologically massive vector field.  All the necessary ingredients for the $\mu l=2$ solution  used  in \cite{Deger:2013yla} are at hand also
in the topologically gauged ${\cal N}=8$ theory discussed in the present paper and we thus expect this kind of $\mu l=2$ solution to exist also here. Note that this $\mu l=2$ solution
is a null-warped one \cite{Deger:2013yla} of the kind that is known to appear in our case for $\mu l=3$.

\acknowledgments

I am grateful to Christian Fronsdal, Ulf Gran, Paul Howe, Sunil Mukhi, Costis Papageorgakis, Per Sundell and Misha Vasiliev  for 
stimulating discussions.  I would also like to thank UG for constructive comments on a preliminary version of the
manuscript. I also thank the Galileo Galilei Institute for Theoretical Physics in Florence for its hospitality 
and the INFN for partial support during the higher spin workshop when the final parts of this paper were written.

%%%%%%%%%%%%%%%%%%%%%%%%%%%%%%%%%%%%%%%%%%%%%%%%%%%%%%%%%

\appendix

\section{Cancelation of  terms in $\de L$ with one or no  $D$}

%%%%%%%%%%%%%%%%%%%%%%%%%%%%%%%%%%%%%%%%%%%%%%%%%%%%%%

Before starting the computation we give our conventions.
We use a mostly plus metric and a Levi-Civita tensor defined by
\beq
\ep^{\mu\nu\rho}:\,\,\,\,\ep^{012}=+1.
\eeq
Then
\beq
\ep^{\mu\nu\rho}\ep_{\tau\nu\rho}=-2e^2\de^{\mu}_{\tau},\,\,\,\,
\ep^{\mu\nu\rho}\ep_{\alpha\beta\rho}=-2e^2\de^{\mu\nu}_{\alpha\beta}.
\eeq
Our gamma matrices satisfy
\beq
\{\ga^{\mu},\ga^{\nu}\}=2g^{\mu\nu},
\eeq
and are chosen such that
\beq
e\ga^{\mu\nu\rho}=\ep^{\mu\nu\rho},\,\,\,e\ga^{\mu\nu}=\ep^{\mu\nu\rho}\ga_{\rho},\,\,\,\,
2e\ga^{\mu}=-\ep^{\mu\nu\rho}\ga_{\nu\rho}.
\eeq

The  lagrangian that we need in the following reads
\beqa
L&=&\tfrac{1}{g}L_{sugra}^{conf}+\tfrac{1}{\al} L_{CS(A)}
-\tfrac{1}{2}eg^{\mu\nu}D_{\mu}X_a^iD_{\nu}X_a^i
+\tfrac{i}{2}e\bar \psi_a\ga^{\mu}D_{\mu}\psi_a\notag\\[1mm]
&+&ieA\bar\chi_{\mu}\Ga^i\ga^{\nu}\ga^{\mu}\psi_a(D_{\nu}X_a^i-\frac{i}{2}\hat A \bar \chi_{\nu}\Ga^i\psi_a)\notag\\[1mm]
&-&iA'\ep^{\mu\nu\rho} \bar \chi_{\mu}\Ga^{ij}\chi_{\nu}(D_{\rho}X_a^i)X_a^j\notag\\[1mm]
&+&iA''\bar f\cdot\ga\Ga^i\psi X_a^i
+iA_{12}\bar f\cdot \chi X^2
+A_{13}eRX^2\notag\\[1mm]
&+&ieA_{14}\bar \psi_a\psi_aX^2+ieA'_{14}\bar \psi_a\psi_b X^i_aX^i_b+ieA_{15}\bar \psi_a\Ga^{ij}\psi_b X^i_aX^j_b\notag\\[1mm]
&+&ie\bar\chi\cdot \ga\Ga^i\psi_a(A_{16}X_a^iX^2+A'_{16}X_a^jX_b^iX_b^j)\notag\\[1mm]
&+&ie\bar\chi\cdot\chi(A_{17}(X^2)^2+A'_{17}(X_a^iX_a^j)(X_b^iX_b^j)) \notag\\[1mm]
&+&ie\ep^{\mu\nu\rho}\bar\chi_{\mu}\ga_{\nu}\chi_{\rho} (A_{18}(X^2)^2+A'_{18}(X_a^iX_a^j)(X_b^iX_b^j))\notag\\[1mm]
&+&eA_{19}(X^2)^3+eA'_{19}(X^2)(X_a^iX_a^j)(X_b^iX_b^j)+eA''_{19}(X_a^iX_a^j)(X_b^jX_b^k)(X_c^kX_c^i),
\eeqa
where all the terms in the first four lines (except $\tfrac{1}{\al} L_{CS(A)}$) were determined in \cite{Gran:2008qx} with the following result:
\beq
\hat A=A,\,\,\,A'=-\tfrac{1}{4},\,\,\,A''=A,\,\,\,A_{12}=\tfrac{1}{4},\,\,\,A_{13}=-\tfrac{1}{16},\,\,\,and \,\,\,A^2=\tfrac{1}{2}.
\eeq
$\tfrac{1}{\al} L_{CS(A)}$  plus the potential  were found in \cite{Gran:2012mg} by various methods. 
This appendix is a continuation of the Noether computation started  in \cite{Gran:2008qx} and supplies the missing details of the presentation in \cite{Gran:2012mg}
where the final result was first presented. Here we also give a more direct argument leading to the normalization of $\tfrac{1}{\al} L_{CS(A)}$ than that given in \cite{Gran:2012mg}. 
The new terms  in $\de\psi$ and $\de B_{\mu}^{ij}$ will be crucial. We  therefore give them explicitly:
\beq
\de \psi_a=\ga^{\mu}\Ga^i\ep_m(D_{\mu}X_a^i-iA \bar\chi_{\mu}\Ga^i\psi_a)+B_5\Ga^i\ep_mX_a^iX^2+B_6\Ga^i\ep_mX_b^iX^j_aX^j_b,\
\eeq
where $A=\pm\tfrac{1}{\sqrt 2}$,
and 
\beqa
\de B_{\mu}^{ij}&=&-\frac{i}{2e}\bar \ep_g\Ga^{ij}\ga_ {\nu}\ga_{\mu}f^{\nu}-\tfrac{ig}{16}\bar\psi_a\ga_{\mu}\Ga^{ijk}\ep_m X_a^k-\tfrac{3ig}{8}\bar\psi_a\ga_{\mu}\Ga^{[i}\ep_m X_a^{j]}
\notag\\[1mm]
&-&\tfrac{ig}{4}\bar \chi_{\mu}\Ga^{k[i}\ep_gX_a^{j]}X_a^k-\tfrac{ig}{32}\bar \chi_{\mu}\Ga^{ij}\ep_gX^2.
\eeqa

Now we add also a variation of the gauge field but without the usual three-algebra structure constant, i.e.,
\beq
\de A_{\mu ab}=2iq\bar\ep_m\ga_{\mu}\Ga^i\psi_{[a}X^i_{b]}+q'i\bar\chi_{\mu}\Ga^{ij}\ep_gX^i_aX^j_b,
\eeq
leading to the following form of the covariant derivative
\beq
D_{\mu}X^i_a=\partial_{\mu} X^i_a+B_{\mu}^{ij}X^j_a+A_{\mu a}{}^b X^i_b.
\eeq

The various kinds of terms with one derivative $D$ that can appear in $\de L$ and need to be canceled are with two fermions
\beq
\ep D\psi X^3,\,\,\,\ep D\chi X^4,
\eeq
and a $D$ together with four fermions
\beq
\ep D\psi \chi\psi,\,\,\,\ep D\psi \chi\chi X,\,\,\ep D\chi \chi^2X^2.
\eeq
The $D^2$ and $D^3$ terms in $\de L$ were dealt with in \cite{Gran:2008qx}. 

%%%%%%%%%%%%%%%%%%%%%%%%%%%%%%%%%%%%%%%%%%%

\subsection{Terms with one $D$ and two fermions: $\ep D\psi X^3$ terms }

%%%%%%%%%%%%%%%%%%%%%%%%%%%%%%%%%%%%%%%%%%%

Starting with the cancelation of $\ep D\psi X^3$ these terms arise from a number of places, namely $\de L_{KG}|_{\de B=\ep\psi X}$, $\de L_{Dirac}|_{\de\psi=\ep X^3}$
and $\de L_{14(Yuk)}|_{\de\psi=\ep DX}$. 

 Adding these should give something that can be canceled by adding a term $\chi\psi X^3$ and vary $\chi$. 
Note that 
$B_5$ and $B_6$ are  obtained  from the computation now to be done.\\
\\
\beqa
&&\de L_{KG} |_{\de B=\ep\psi X, \de A=\ep\psi X}\notag\\[1mm]
&&=-e(D^{\mu}X_a^i)X_a^j(-\tfrac{ig}{16}\bar\psi_b\ga_{\mu}\Ga^{ijk}\ep_m X_b^k-\tfrac{3ig}{8}\bar\psi_b\ga_{\mu}\Ga^{[i}\ep_m X_b^{j]})\notag\\[1mm]
&&-iqe(D^{\mu}X_a^i)X_b^i\bar\ep_m\ga_{\mu}\Ga^j( \psi_aX^j_b- \psi_bX^j_a),
\eeqa
where we see that the first term needs to be canceled by the Yukawa term containing $\Ga^{ij}$ and the other can be written with the antisymmetry written out and
with an index $b$ on the spinor and $i$ on the $\Ga$ in all terms:
\beqa
&&\de L_{4-KG} |_{\de B=\ep\psi X, \de A=\ep\psi X}\notag\\[1mm]
&=&-e(D^{\mu}X_a^i)X_a^j(-\tfrac{ig}{16}\bar\psi_b\ga_{\mu}\Ga^{ijk}\ep_m X_b^k-\tfrac{3ig}{16}\bar\psi_b\ga_{\mu}\Ga^{i}\ep_m X_b^{j}+\tfrac{3ig}{16}\bar\psi_b\ga_{\mu}\Ga^{j}\ep_m X_b^{i})\notag\\[1mm]
&&-iqe(D^{\mu}X_b^j)X_a^j\bar\ep_m\ga_{\mu}\Ga^i \psi_bX^i_a+iqe(D^{\mu}X_a^j)X_b^j\bar\ep_m\ga_{\mu}\Ga^i \psi_bX^i_a.
\eeqa
Next we derive the contribution from $\de L_{5(Dirac)}|_{\de\psi=\ep X^3}$:
\beqa
&&\de L_{5-Dirac}|_{\de\psi=\ep X^3}\notag\\[1mm]
&=&ie\bar\psi_b\Ga^i\ga^{\mu}\tilde D_{\mu}\ep_m(B_5X_b^iX^2+B_6X_b^jX_a^iX_a^j)\notag\\[1mm]
&+&ie\bar\psi_b\Ga^i\ga^{\mu}\ep_m\tilde D_{\mu}(B_5X_b^iX^2+B_6X_b^jX_a^iX_a^j),
\eeqa
and from $\de L_{Yuk}|_{\de\psi=\ep DX}$ we get:
\beqa
&&\de L_{Yuk}|_{\de\psi=\ep DX}\notag\\[1mm]
&=&2ieA_{14}\bar\psi_b\Ga^i\ga^{\mu}\ep_m(\tilde D_{\mu}X_b^i)X^2+2ieA'_{14}\bar\psi_b\Ga^i\ga^{\mu}\ep_m(\tilde D_{\mu}X_a^i)X^j_aX^j_b\notag\\[1mm]
&+&2ieA_{15}\bar\psi_b\Ga^{jk}\Ga^i\ga^{\mu}\ep_m(\tilde D_{\mu}X_a^i)X_b^jX_a^k\notag\\[1mm]
&=&2ieA_{14}\bar\psi_b\Ga^i\ga^{\mu}\ep_m(\tilde D_{\mu}X_b^i)X^2+2ieA'_{14}\bar\psi_b\Ga^i\ga^{\mu}\ep_m(\tilde D_{\mu}X_a^i)X^j_aX^j_b\notag\\[1mm]
&+&2ieA_{15}\bar\psi_b\Ga^{ijk}\ga^{\mu}\ep_m(\tilde D_{\mu}X_a^i)X_b^jX_a^k\notag\\[1mm]
&+&ieA_{15}\bar\psi_b\Ga^{i}\ga^{\mu}\ep_m(\tilde D_{\mu}X^2)X_b^i-2ieA_{15}\bar\psi_b\Ga^{i}\ga^{\mu}\ep_m(\tilde D_{\mu}X_a^j)X_b^jX_a^i.
\eeqa

Since we are avoiding derivatives on $\psi$ we must cancel terms as they are without integrations by part. Then all terms except the $D\ep$ must cancel
directly. The first terms to cancel are the $\Ga^{ijk}$ terms giving
\beq
2A_{15}-\tfrac{g}{16}=0.
\eeq

 Then from the cancelation of $\bar\psi...\ep_mX^2\tilde D_{\mu}X_b^i$  and  $\bar\psi...\ep_mX_b^i\tilde D_{\mu}X^2$ we get  
\beq
B_5+2A_{14}=0,\,\,\,\,
B_5+A_{15}=0,
\eeq
which implies
\beq
B_5=-\tfrac{g}{32},\,\,\,B_5=-2A_{14},\,\,\,A_{15}=\tfrac{g}{32}.
\eeq

Looking now at the terms $(D^{\mu}X_a^j)X_a^iX_b^j$ and $(D^{\mu}X_a^i)X_a^jX_b^j$ we find cancelation for
\beq
-\tfrac{3g}{16}+B_6-2A_{15}+q=0,
\eeq
and
\beq
\tfrac{3g}{16}+B_6+2A'_{14}=0,
\eeq
and for the last kind of such terms $(D^{\mu}X_b^j)X_a^iX_a^j$:
\beq
B_6-q=0,
\eeq
%In order to deal with this last result, which cannot be correct(????), we use the trick of cycling  indices in the expression
%\beq
%\bar\psi_b\Ga^i\ga^{\mu}\ep_m(D^{\mu}X_b^j)X_a^iX_a^j
%\eeq
%As shown in the Appendix by replacing $\Ga^i$ by $\tfrac{1}{7!}\ep^{ik_1.....k_7}\Ga^{k_1.....k_7}$ and cycling the nine indices
%$ijk_1......k_7$ on $\Ga^{k_1.....k_7}(D^{\mu}X_b^j)X_a^i$ one gets the result
%\beq
%\bar\psi_b\Ga^i\ga^{\mu}(D^{\mu}X_b^j)X_a^iX_a^j=\bar\psi_b\Ga^i\ga^{\mu}(D^{\mu}X_b^i)X^2
%\eeq
%so that this actually contributes to a cancelation of $X^2$ terms discussed above.
giving the result
\beq
B_6=q=\tfrac{g}{8},\,\,\,A'_{14}=-\tfrac{5g}{32},\,\,\,using \,\,\,A_{15}=\tfrac{g}{32}.
\eeq

Finally to cancel the $\tilde D_{\mu}\ep$ term we must add
\beq
ie\bar\chi\cdot \ga\Ga^i\psi_b (A_{16}X_b^iX^2+A'_{16}X_b^jX_a^jX_a^i)
\eeq
with
\beq
A_{16}=-AB_5\,\,\,and\,\,\,A'_{16}=-AB_6
\eeq

%%%%%%%%%%%%%%%%%%%%%%%%%%%%%%%%%%%%%%%%%%%

\subsection{Terms with one $D$ and two fermions:  $\ep D\chi X^4$ terms }

%%%%%%%%%%%%%%%%%%%%%%%%%%%%%%%%%%%%%%%%%%%

These  come from the following variations
\beqa
\de L_{4(KG)}|_{\de B=\ep\chi X, \de A=\ep\chi X}&=&e(\tilde D_{\mu}X_a^i)X_a^j(\tfrac{ig}{4}\bar\chi^{\mu}\Ga^{k[i}\ep_gX_b^{j]}X_b^k+\tfrac{ig}{32}\bar\chi^{\mu}\Ga^{ij}\ep_gX^2)\notag\\[1mm]
&&-q'ie(\tilde D_{\mu}X_a^i)X_b^i\bar\chi_{\mu}\Ga^{jk}\ep_gX^j_aX^k_b,
\eeqa
\beqa
&&\de L_{9(SC)}|_{\de\psi=\ep X^3}\notag\\[1mm]
&&=ieAB_5\bar\chi_{\nu}\Ga^i\Ga^j\ga^{\mu}\ga^{\nu}\ep_m(\tilde D_{\mu}X_a^i)X_a^jX^2\notag\\[1mm]
&&+ieAB_6\bar\chi_{\nu}\Ga^i\Ga^j\ga^{\mu}\ga^{\nu}\ep_m(\tilde D_{\mu}X_a^i)X_b^jX^k_aX^k_b,
\eeqa
\beqa
&&\de L_{10}|_{\de\psi=\ep X^3}\notag\\[1mm]
&&=iA''B_5\bar f^{\mu}\ga_{\mu}\Ga^i\Ga^j\ep_mX_a^jX^2X_a^i+iA''B_6\bar f^{\mu}\ga_{\mu}\Ga^i\Ga^j\ep_mX_b^jX_a^kX_b^kX_a^i\notag\\[1mm]
&&=iA''B_5\bar f^{\mu}\ga_{\mu}\ep_m(X^2)^2+iA''B_6\bar f^{\mu}\ga_{\mu}\ep_mX_a^iX_a^jX_b^iX_b^j,
\eeqa
since the $\Ga^{ij}$ term vanishes! Next term is
\beqa
&&\de L_{16}|_{\de\psi=\ep \tilde DX}=ie\bar\chi\cdot \ga\Ga^i\de\psi_b (A_{16}X_b^iX^2+A'_{16}X_b^jX_a^jX_a^i)\notag\\[1mm]
&&=ie\bar\chi\cdot \ga\Ga^i\Ga^k\ga^{\mu}\ep_m(D_{\mu}X^k_a )(A_{16}X_a^iX^2+A'_{16}X_a^jX_b^jX_b^i).
\eeqa
Here there will be a nice test of the coefficients so far since all $\Ga^{ij}$ terms must cancel when summing up the expressions above. The reason is that no
$\chi^2X^4$ terms can be written down with $\Ga^{ij}$ matrices.

We now have all the contributions and can start to require cancelations from susy. First, the $\Ga^{ij}$ matrix terms give for the $X^2$ terms, using also the relation for $B_4$,
\beq
\tfrac{g}{32}g^{\mu\nu}\ep_g+AB_5\ga^{\mu}\ga^{\nu}\ep_m-A_{16}\ga^{\nu}\ga^{\mu}\ep_m=0,
\eeq
which means
\beq
\tfrac{g}{32}g^{\mu\nu}\ep_g+A^2B_5\ga^{\mu}\ga^{\nu}\ep_g-AA_{16}\ga^{\nu}\ga^{\mu}\ep_g=0,
\eeq
giving for the $\ga^{\mu\nu}$ terms
\beq
\tfrac{1}{2}B_5+AA_{16}=0,
\eeq
and for the $g^{\mu\nu}$ terms
\beq
\tfrac{g}{32}+\tfrac{1}{2}B_5-AA_{16}=0.
\eeq
Adding and  subtracting them give the following two equations 
\beq
B_5=-\tfrac{g}{32},\,\,\,\,
AA_{16}=\tfrac{g}{64}.
\eeq

Next we turn to the $\Ga^{ij}$ matrix terms give for the non-$X^2$ terms
\beqa
&&\tfrac{ig}{8}e(\tilde D_{\mu}X_a^i)X_a^j(\bar\chi^{\mu}\Ga^{ki}\ep_gX_b^{j}X_b^k-\bar\chi^{\mu}\Ga^{kj}\ep_gX_b^{i}X_b^k)\notag\\[1mm]
&&-q'ie(\tilde D_{\mu}X_a^i)X_a^j\bar\chi_{\mu}\Ga^{jk}\ep_gX^i_bX^k_b\notag\\[1mm]
&&+ieAB_6\bar\chi_{\nu}\Ga^{ij}\ga^{\mu}\ga^{\nu}\ep_m(\tilde D_{\mu}X_a^i)X_b^jX^k_aX^k_b\notag\\[1mm]
&&+ieA'_{16}\bar\chi\cdot \ga\Ga^{ik}\ga^{\mu}\ep_m(D_{\mu}X^k_a )X_a^jX_b^jX_b^i=0
\eeqa
Changing indices to get the same factor of $(DX)X$ and then dropping it gives
\beqa
&&-\tfrac{g}{8}\bar\chi^{\mu}\Ga^{ik}\ep_gX_b^{j}X_b^k+\tfrac{g}{8}\bar\chi^{\mu}\Ga^{jk}\ep_gX_b^{i}X_b^k
-q'\bar\chi_{\mu}\Ga^{jk}\ep_gX^i_bX^k_b\notag\\[1mm]
&&+AB_6\bar\chi_{\nu}\Ga^{ik}\ga^{\mu}\ga^{\nu}\ep_mX_b^kX^j_b
-A'_{16}\bar\chi\cdot \ga\Ga^{ik}\ga^{\mu}\ep_mX_b^jX_b^k=0.
\eeqa
The $\ga$-terms must give rise to an anticommutator which means that
\beq
A'_{16}=-AB_6,
\eeq
and then the whole equation becomes
\beqa
&&-\tfrac{g}{8}\bar\chi^{\mu}\Ga^{ik}\ep_gX_b^{j}X_b^k+\tfrac{g}{8}\bar\chi^{\mu}\Ga^{jk}\ep_gX_b^{i}X_b^k
-q'\bar\chi_{\mu}\Ga^{jk}\ep_gX^i_bX^k_b\notag\\[1mm]
&&+2AB_6\bar\chi_{\nu}\Ga^{ik}\ep_mX_b^kX^j_b=0.
\eeqa
Using that $\ep_m=A\ep_g$ then gives
\beqa
&&-\tfrac{g}{8}\Ga^{ik}X_b^{j}X_b^k+\tfrac{g}{8}\Ga^{jk}X_b^{i}X_b^k
-q'\Ga^{jk}X^i_bX^k_b+2A^2B_6\Ga^{ik}X_b^kX^j_b=0,
\eeqa
implying
\beq
q'=\tfrac{g}{8},\,\,\,2A^2B_6=\tfrac{g}{8}\,\,\,or \,\,\,B_6=\tfrac{g}{8}.
\eeq

Now we check the remaining terms, i.e., those without $\Ga$-matrices
\beqa
&&\tfrac{ie}{4}AB_5\bar\chi_{\nu}\ga^{\mu}\ga^{\nu}\ep_m\tilde D_{\mu}(X^2)^2
+\tfrac{ie}{4}AB_6\bar\chi_{\nu}\ga^{\mu}\ga^{\nu}\ep_m\tilde D_{\mu}(X^i_aX^i_bX^j_aX^j_b)\notag\\[1mm]
&&+iA''B_5\bar f^{\mu}\ga_{\mu}\ep_m(X^2)^2+iA''B_6\bar f^{\mu}\ga_{\mu}\ep_mX^i_aX^i_bX^j_aX^j_b\notag\\[1mm]
&&+\tfrac{ie}{4}A_{16}\bar\chi_{\nu}\ga^{\nu}\ga^{\mu}\ep_m\tilde D_{\mu}(X^2)^2
+\tfrac{ie}{4}A'_{16}\bar\chi_{\nu}\ga^{\nu}\ga^{\mu}\ep_m\tilde D_{\mu}(X^i_aX^i_bX^j_aX^j_b),
\eeqa
where the last two terms come from the above variation of $\tilde\om$ in the $RX^2$ term. Note that the very last term then cancels the second term!

Then with $e\ga^{\mu\nu}=\ep^{\mu\nu\rho}\ga_{\rho}$ the $X^2$ terms  containing $f$ become (the rest of the terms work the same way)
\beqa
&&+iA''B_5\bar f^{\mu}\ga_{\mu}\ep_m(X^2)^2=\tfrac{i}{2}A''B_5\ep^{\mu\nu\rho}\tilde D_{\nu}\bar\chi_{\rho}\ga_{\mu}\ep_m(X^2)^2\notag\\[1mm]
&&=-\tfrac{i}{2}A''B_5\ep^{\mu\nu\rho}\bar\chi_{\rho}\ga_{\mu}(\tilde D_{\nu}\ep_m)(X^2)^2-\tfrac{i}{2}A''B_5\ep^{\mu\nu\rho}\bar\chi_{\rho}\ga_{\mu}\ep_m(\tilde D_{\nu}(X^2)^2)+contortion\notag\\[1mm]
&&=\tfrac{ie}{2}A''B_5\bar\chi_{\mu}\ga^{\mu\nu}(\tilde D_{\nu}\ep_m)(X^2)^2-\tfrac{ie}{2}A''B_5\bar\chi_{\mu}\ga^{\mu\nu}\ep_m(\tilde D_{\nu}(X^2)^2)+contortion,
\eeqa
after an integration by parts. 

We can now collect and cancel the $D(X^2)^2g^{\mu\nu}\ep_m$ terms: 
%\beq
%\tfrac{1}{4}AB_5+\tfrac{1}{4}A_{16}=0,
%\eeq
%or
\beq
AA_{16}=-\tfrac{1}{2}B_5,
\eeq
while the antisymmetric part implies
\beq
-\tfrac{1}{4}AB_5+\tfrac{1}{4}A_{16}+\tfrac{1}{2}A''B_5=0,
\eeq
which just means that the previous relation is obtained  once  again.
%\beq
%AA_{16}=-\tfrac{1}{2}B_5.
%\eeq
%

Add now term 18 in $L$
\beq
L_{18}=ieA_{18}\ep^{\mu\nu\rho}\bar\chi_{\mu}\ga_{\nu}\chi_{\rho} (X^2)^2,
\eeq
which varies into
\beq
\de L_{18}=2ieA_{18}\ep^{\mu\nu\rho}\tilde D_{\mu}\bar\ep_g\ga_{\nu}\chi_{\rho} (X^2)^2-
4eA_{18}\ep^{\mu\nu\rho}\bar\chi_{\mu}\ga_{\nu}\chi_{\rho} (X^2)X_a^i\bar\ep_m\Ga^i\psi_a.
\eeq
Thus if 
\beq
2A_{18}\ep_g=\tfrac{1}{2}AB_5\ep_m,
\eeq
the one-derivative terms cancel so (sing $2A^2=1$)
%\beq
%2A_{18}=\tfrac{1}{2}A^2B_5,
%\eeq
%or
\beq
A_{18}=\tfrac{1}{8}B_5.
\eeq
Since the other terms work the same way the full new term in L is
\beq
L_{18}=ieA_{18}\ep^{\mu\nu\rho}\bar\chi_{\mu}\ga_{\nu}\chi_{\rho} (X^2)^2+ieA'_{18}\ep^{\mu\nu\rho}\bar\chi_{\mu}\ga_{\nu}\chi_{\rho} X^i_aX^i_bX^j_aX^j_b,
\eeq
with 
\beq
A_{18}=\tfrac{1}{8}B_5\,\,\,and\,\,\,A'_{18}=\tfrac{1}{8}B_6.
\eeq

Note that no $\chi^2X^2$ term without the Levi-Civita tensor is not needed just as in ABJM. With the obtained values we see that
\beq
A_{18}=-\tfrac{g}{256}\,\,\,and\,\,\,A'_{18}=\tfrac{g}{64}
\eeq

%%%%%%%%%%%%%%%%%%%%%%%%%%%%%%%%%%%%%%%%%%%

\subsection{The normalization of the CS term for the  gauge field $A_{\mu}^{ab}$}

%%%%%%%%%%%%%%%%%%%%%%%%%%%%%%%%%%%%%%%%%%%

After having determined the coefficients $q$ and $q'$ in the variation $\de A_{\mu}^{ab}$ we must now 
return to the question of the corresponding CS term appearing in $L$ and its normalization  in terms of the parameter
$\al$. We will trace the places in the previous derivation of the lagrangian where the field strength $F_{\mu\nu}^{ab}$ appears
simply by looking for where $G_{\mu\nu}^{ab}$ appears as a result of evaluating the commutator of two covariant derivatives 
acting on $X^i_a$. Note that this computation also arises acting on the supersymmetry parameter in some cases but then 
$F_{\mu\nu}^{ab}$ will not appear since the susy parameter is inert under gauge symmetry.

There are two places where $F_{\mu\nu}^{ab}$ appears: in the variation of the Dirac kinetic term giving
\beq
\tfrac{i}{2}\ep^{\mu\nu\rho}\bar \psi_a\ga_{\rho}\Ga^i\ep_m(G_{\mu\nu}^{ij}X^j_a+F_{\mu\nu}^{ab}X^i_b),
\eeq
and from the variation of the term denoted $L'$
\beq
iA'\ep^{\mu\nu\rho}\bar \chi_{\mu}\Ga^{ik}\ep_g(G_{\mu\nu}^{ij}X^j_a+F_{\mu\nu}^{ab}X^i_b)X^k_a.
\eeq

These contributions to $\de L$ must be cancelled by adding terms to the variation of the gauge fields using
\beq
\de L_{CS(B,A)}=\tfrac{1}{g}\ep^{\mu\nu\rho}\de B_{\mu}^{ij}|_{new}G_{\nu\rho}^{ij}+\tfrac{1}{2\al}\ep^{\mu\nu\rho}\de A_{\mu}^{ab}|_{new}F_{\nu\rho}^{ab}.
\eeq
For the R-symmetry terms this implies
\beq
B_2=-\tfrac{g}{2},\,\,\,B_3=gA',
\eeq
as we already have seen. However, for the gauge field $ A_{\mu}^{ab}$ the results are new and read
\beq
\al=-2q,\,\,\,2\al A'=q',
\eeq
which must give the same answer for $\al$.  Inserting $q=q'=\tfrac{g}{8}$ and $A'=-\tfrac{1}{4}$
we find that this is indeed the case:
\beq
\al=-\tfrac{g}{4}.
\eeq

%%%%%%%%%%%%%%%%%%%%%%%%%%%%%%%%%%%%%%%%%%%

%\subsubsection{Terms with one $D$ and four fermions}

%%%%%%%%%%%%%%%%%%%%%%%%%%%%%%%%%%%%%%%%%%%

%%%%%%%%%%%%%%%%%%%%%%%%%%%%%%%%%%%%%%%%%%%

\subsection{Cancellation of terms with no $D$ and two fermions}

%%%%%%%%%%%%%%%%%%%%%%%%%%%%%%%%%%%%%%%%%%%

Here we concentrate on the cancellations that will lead us to the form of the potential.
%However, at the end we include one of the many cancellations that requires fierzing 
%to see that we have obtained sensible results so far. As mentioned the fact that the results
%obtained here are correct and part of a consistent theory follows from the results presented in
%\cite{Gran:2012mg} which we refer the interested reader to.

%%%%%%%%%%%%%%%%%%%%%%%%%%%%%%%%%%%%%%%%%%%

%\subsubsection{Terms with no $D$ and two fermions}

%%%%%%%%%%%%%%%%%%%%%%%%%%%%%%%%%%%%%%%%%%%

Start by varying the $X^6$ potential
\beq
L_{X^6}=eA_{19}(X^2)^3+eA'_{19}(X^2)(X_a^iX_a^j)(X_b^iX_b^j)+eA''_{19}(X_a^iX_a^j)(X_b^jX_b^k)(X_c^kX_c^i).
\eeq
We find  that varying this term gives
\beqa
&&\de L_{X^6}
=ie(\bar\ep_g\ga^{\mu}\chi_{\mu})(A_{19}(X^2)^3+A'_{19}(X^2)(X_a^iX_a^j)(X_b^iX_b^j)\notag\\[1mm]
&&+A''_{19}(X_a^iX_a^j)(X_b^jX_b^k)(X_c^kX_c^i))\notag\\[1mm]
&&+ieA_{19}6(X^2)^2X_a^i\bar\ep_m\Ga^i\psi_a\notag\\[1mm]
&&+ieA'_{19}(2X_c^k\bar\ep_m\Ga^k\psi_c(X_a^iX_a^j)(X_b^iX_b^j)+4X^2(X_a^i\bar\ep_m\Ga^j\psi_a)(X_b^iX_b^j))\notag\\[1mm]
&&+ieA''_{19}(6\bar\ep_m\Ga^i\psi_aX_a^j)(X_b^jX_b^k)(X_c^kX_c^i).
\eeqa

From the $\chi$ terms we can obtain uniquely the $A_{16}$ coefficients in front of the $\chi\psi X^3$ terms using $\de \psi=\ep X^3$.
This variation reads
\beqa
&&\de L_{\chi\psi X^3}|_{\de \psi|_{\ep X^3}}=ie\bar\chi\cdot \ga\Ga^i(\de \psi_a)|_{\ep X^3}(A_{16}X_a^iX^2+A'_{16}X_a^jX_b^iX_b^j)\notag\\[1mm]
&&=ie\bar\chi\cdot \ga\Ga^i\Ga^k\ep_m(B_5X_a^kX^l_cX^l_c+B_6X_c^kX^l_aX^l_c  )(A_{16}X_a^iX^2+A'_{16}X_a^jX_b^iX_b^j).
\eeqa
Here all $\Ga^{ik}$ terms vanish since all expressions in terms of six scalars are symmetric in two free $ik$ indices. Thus the above becomes
\beqa
&&\de L_{\chi\psi X^3}|_{\de \psi|_{\ep X^3}}=ie\bar\chi\cdot \ga\ep_m(B_5X_a^iX^l_cX^l_c+B_6X_c^iX^l_aX^l_c  )(A_{16}X_a^iX^2+A'_{16}X_a^jX_b^iX_b^j)\notag\\[1mm]
&&=-ie\bar\ep_m\ga\cdot \chi(B_5A_{16}(X^2)^3+(B_5A'_{16}+B_6A_{16})X^2X^{ij}X^{ij}+B_6A'_{16}X^{ij}X^{jk}X^{ki}).
\eeqa
Thus the  cancelation of these terms gives the relations
\beq
A_{19}=B_5(AA_{16}),\,\,\,A'_{19}=B_5(AA'_{16})+B_6(AA_{16}),\,\,\,A''_{19}=B_6(AA'_{16})
\eeq

Now recall
\beqa
&&\de L_{5(Dirac)}|_{\de\psi=\ep X^3}\notag\\[1mm]
&=&ie\bar\psi_b\Ga^i\ga^{\mu}\tilde D_{\mu}\ep_m(B_5X_b^iX^2+B_6X_b^jX_a^iX_a^j)\notag\\[1mm]
&+&ie\bar\psi_b\Ga^i\ga^{\mu}\ep_m\tilde D_{\mu}(B_5X_b^iX^2+B_6X_b^jX_a^iX_a^j),
\eeqa
where only the $D\ep$ terms remain to be canceled which is done by the term 
\beq
L_{\chi\psi X^3}=ie\bar\chi\cdot \ga\Ga^i\psi_a(A_{16}X_a^iX^2+A'_{16}X_a^jX_b^iX_b^j).
\eeq
The $\de\chi_{\mu}=D_{\mu}\ep_g$ variation gives
\beqa
\de L_{\chi\psi X^3}|_{ \de\chi_{\mu}=D_{\mu}\ep_g}&&=ieD_{\mu}\bar\ep_g \ga_{\mu}\Ga^i\psi_a(A_{16}X_a^iX^2+A'_{16}X_a^jX_b^iX_b^j)\notag\\[1mm]
&&=ie\bar\psi_a \ga_{\mu}\Ga^iD_{\mu}\ep_g(A_{16}X_a^iX^2+A'_{16}X_a^jX_b^iX_b^j).
\eeqa
Cancelation implies
\beq
AB_5=-A_{16},\,\,\,AB_6=-A'_{16}.
\eeq

Hence we know the six order potential:
\beqa
&&A_{19}=-A_{16}^2=-(AB_5)^2=-\tfrac{g^2}{2\cdot 32\cdot 32},\notag\\[1mm]
&&A'_{19}=-2A_{16}A'_{16}=-2A^2B_5B_6=\tfrac{g^2}{8\cdot 32},\notag\\[1mm]
&&A''_{19}=-(A'_{16})^{2}=-(AB_6)^2
=-\tfrac{g^2}{2\cdot 8\cdot 8}.
\eeqa

With a potential the theory should have an $AdS$ vacuum that puts the theory at a chiral point. If we set the VEV $< X >=v$ 
 we find that the potential gives
\beq
L_{X^6}(v)=(A_{19}+A'_{19}+A''_{19})v^6.
\eeq
Adding the gravitational CS term and the $X^2R$ term evaluated at the VEV we get
\beq
L_{AdS}=L_{CS(\om)}-\tfrac{v^2e}{16}R+L_{X^6}(v)
\eeq
We should compare this to Li, Song and Strominger (LSS) for the chiral point but with TMG signs in the lagrangian:
\beq
L_{LSS}=\frac{1}{\kappa^2}(\frac{1}{\mu}L_{CS(\om)}-e(R-2\Lambda)).
\eeq
Thus $\mu=\tfrac{1}{\kappa^2}$  and $v^2=\tfrac{16}{\kappa^2}$. The chiral point condition is $\mu l=1$ where $l$ is defined 
by means of the cosmological constant as $\Lambda=-\tfrac{1}{l^2}$. This implies that, to end up a chiral point, the potential must  satisfy
\beq
\tfrac{1}{e}L_{X^6}(v)=\tfrac{2\Lambda}{\kappa^2}=-\tfrac{2}{\kappa^2 l^2}=-\tfrac{2\mu^2}{\kappa^2}=-\tfrac{2}{\kappa^6}=-\tfrac{2v^6}{16^3}.
\eeq

Thus we see that for the theory to be at the chiral point we must require
\beq
A_{19}+A'_{19}+A''_{19}=-\tfrac{2}{16^3}=-\tfrac{1}{2048},
\eeq
(which strangely enough happens to be exactly $A_{19}$ above!).

Next we consider the variation of the Yukawa terms that connect to the variation of the $X^6$ potential above
\beq
L_{Yuk}=ieA_{14}\bar \psi_a\psi_aX^2+ieA'_{14}\bar \psi_a\psi_b X^i_aX^i_b\\
+ieA_{15}\bar \psi_a\Ga^{ij}\psi_b X^i_aX^j_b\\
\eeq
Vary this using the $\psi=\ep X^3$ expression
\beq
\de \psi_a|_{\ep X^3}=B_5\Ga^k\ep_mX_a^kX^l_bX^l_b+B_6\Ga^k\ep_mX_b^kX^l_aX^l_b\\
\eeq
We get
\beqa
\de L_{Yuk}&&=2ieA_{14}\bar \psi_a\de \psi_aX^2+2ieA'_{14}\bar \psi_a\de \psi_b X^i_aX^i_b
+2ieA_{15}\bar \psi_a\Ga^{ij}\de \psi_b X^i_aX^j_b\notag\\[1mm]
&&=2ieA_{14}\bar \psi_a( B_5\Ga^k\ep_mX_a^kX^2+B_6\Ga^k\ep_mX_b^kX^l_aX^l_b)X^2\notag\\[1mm]
&&+2ieA'_{14}\bar \psi_a(B_5\Ga^k\ep_mX_b^kX^2+B_6\Ga^k\ep_mX_c^kX^l_bX^l_c ) X^i_aX^i_b\notag\\[1mm]
&&+2ieA_{15}\bar \psi_a\Ga^{ij}(B_5\Ga^k\ep_mX_b^kX^2+B_6\Ga^k\ep_mX_c^kX^l_bX^l_c ) X^i_aX^j_b
\eeqa

From the conformal variation of the spin 3/2 field in the 16'th term in L we get
\beqa
(\de L_{16}+\de L_{16'})|_{\chi=\ga\ep X^2}=3ieB_7\bar\ep_m\Ga^i\psi_a(A_{16}X_a^i(X^2)^2+A'_{16}X_a^jX_b^iX_b^jX^2)
\eeqa
Cancelation gives the relations
\beqa
(X^2)^2X^i:&&6A_{19}=2B_5(A_{14}+A_{15})\notag\\[1mm]
(X^{jk}X^{jk})X^i:&&2A'_{19}=2B_6A_{15}\notag\\[1mm]
X^2X^{ji}X^i:&&4A'_{19}=2B_5(A'_{14}-A_{15})+2B_6A_{14}\notag\\[1mm]
X^{kj}X^{ji}X^i:&&6A''_{19}=2B_6(A'_{14}-A_{15}).
\eeqa
Inserting the values of the various parameters on the right hand sides as derived previously we confirm the values
of $A_{19}, A'_{19}, A''_{19}$ found above.

\end{document}